\documentclass[sigconf]{acmart}

\usepackage[colorinlistoftodos]{todonotes}

\usepackage{supertabular}   

\usepackage{enumitem}

\AtBeginDocument{%
	}

\setcopyright{none}

\usepackage{url}
\usepackage{hyperref}
\usepackage[hyphenbreaks]{breakurl}

\copyrightyear{2024}
\acmYear{2024}
\setcopyright{rightsretained}
\acmConference[FAccT '24]{The 2024 ACM Conference on Fairness, Accountability, and Transparency}{June 3--6, 2024}{Rio de Janeiro, Brazil}
\acmBooktitle{The 2024 ACM Conference on Fairness, Accountability, and Transparency (FAccT '24), June 3--6, 2024, Rio de Janeiro, Brazil}\acmDOI{10.1145/3630106.3658991}
\acmISBN{979-8-4007-0450-5/24/06}

\begin{document}
	
	\title{\textbf{Evidence of What, for Whom? The Socially Contested Role of Algorithmic Bias in a Predictive Policing Tool}}
	
	\author{Marta Ziosi}
 \authornote{Both authors contributed equally to this work.}
 \affiliation{%
		\institution{Oxford University}
		\streetaddress{Oxford}
		\city{Oxford}
		\country{UK}
		\postcode{12345}
	}
\email{marta.ziosi@sant.ox.ac.uk}
    \author{Dasha Pruss}
    \authornotemark[1]
    
	\affiliation{%
		\institution{Harvard University}
		\streetaddress{Harvard}
		\city{Cambridge}
		\country{USA}
		\postcode{02138}
	}
\email{dpruss@fas.harvard.edu}

	\begin{abstract}
		 This paper presents a critical, qualitative study of the social role of algorithmic bias in the context of the Chicago crime prediction algorithm, a predictive policing tool that forecasts when and where in the city crime is most likely to occur. Through interviews with 18 Chicago-area community organizations, academic researchers, and public sector actors, we show that stakeholders from different groups articulate diverse problem diagnoses of the tool’s algorithmic bias, strategically using it as evidence to advance criminal justice interventions that align with stakeholders' positionality and political ends. Drawing inspiration from Catherine D'Ignazio’s taxonomy of ``refusing and using'' data, we find that stakeholders use evidence of algorithmic bias to reform the policies around police patrol allocation; reject algorithm-based policing interventions; reframe crime as a structural rather than interpersonal problem; reveal data on authority figures in an effort to subvert their power; repair and heal families and communities; and, in the case of more powerful actors, to reaffirm their own authority or existing power structures. We identify the implicit assumptions and scope of these varied uses of algorithmic bias as evidence, showing that they require different (and sometimes conflicting) values about policing and AI. This divergence reflects long-standing tensions in the criminal justice reform landscape between the values of liberation and healing often centered by system-impacted communities and the values of surveillance and deterrence often instantiated in data-driven reform measures. We advocate for centering the interests and experiential knowledge of communities impacted by incarceration to ensure that evidence of algorithmic bias can serve as a device to challenge the status quo.

	\end{abstract}

\ccsdesc[500]{Human-centered computing~Empirical studies in HCI}
 
	\keywords{criminal justice; predictive policing; algorithmic bias; qualitative}
	
	
	\maketitle
	
	\section{Introduction}

Over the past two decades, public scrutiny and outrage toward policing scandals has led some of the largest police departments in the US to turn to ``data-driven'' policing as a reform and public relations measure \cite{brayne_predict_2020}. Place-based predictive policing tools, which forecast crime ``hotspots'' based on past crime incidence data, have become an increasingly popular and controversial strategy for addressing a range of issues in policing, from resource limitations to unconscious biases in decision-making \cite{selbst2017disparate}. 
 
 Proponents laud predictive policing as an evidence-based policing reform that promises more objective decision-making \cite{kennedy_risk-based_2018, pearsall2010predictive, porter_trust_1995, espeland_accountability_2007}, while critics argue that it exacerbates the very issues it is meant to address \cite{harcourt_against_2008, shapiro_predictive_2019, richardson_dirty_2019}. Algorithmic bias -- the disparate treatment of or impact on protected groups by an algorithm -- has been at the center of the discourse about predictive policing, including concerns about the algorithms' reliance on ``dirty data'' and the possibility of feedback loops in officers' arresting behavior \cite{lum_predict_2016, richardson_dirty_2019, ensign_runaway_2018, selbst2017disparate}.

Standard algorithmic fairness approaches have focused on identifying, removing, or managing statistical biases \cite{fazelpour_algorithmic_2021}, with strategies ranging from pre-, during and post-deployment strategies to preventing models from learning or outputting biases  \cite{ensign_runaway_2018, fahse_managing_2021, yeung_identifying_2021, zhang_mitigating_2018}. However, many critics have pointed out that verifying that an algorithm is correctly `calibrated' is not the right question to ask when underlying data reflect structural injustices \cite{balayn2021beyond, hoffmann2019fairness}. Indeed, the bias exhibited by predictive policing algorithms is considered by many scholars to be an inevitable artifact of higher police presence in historically marginalized communities \cite{lum_predict_2016, zilka_progression_2023}, a reflection of the structural race- and class-biased patterns of the criminal legal system and an unequal society more broadly, rather than a technical ``glitch'' \cite{benjamin_race_2019}. In response, some researchers have called for more substantive notions of algorithmic fairness that take algorithms' objectives and sociotechnical contexts into account \cite{barabas_beyond_2019, green_algorithmic_2020, green_escaping_2022}. Others have focused on providing affected communities with alternative tools to respond to the deeper societal issues indicated by algorithmic bias, such as the use of civilian dashcam 
data to detect disparities in police deployments \cite{franchi_detecting_2023}. In short, there has been widespread disagreement about how to respond to evidence of algorithmic bias.

This paper presents a critical, qualitative study on the socially contested role of algorithmic bias in the context of the Chicago crime prediction algorithm \cite{rotaru_event-level_2022}, a predictive policing tool that predicts the geographic regions in Chicago in which crime is most likely to occur. The algorithm's developers demonstrate that it exhibits ``enforcement bias'' -- that is, the uneven or disproportionate patrolling, arrests, or other actions by police toward certain neighborhoods. 

Through interviews with Chicago-area academic researchers, 
community organizations, 
and public sector employees, 
we find that evidence of algorithmic bias has the potential to challenge as much as to affirm the status quo. We show that stakeholders from different groups articulate different problem diagnoses of the algorithm's bias, strategically using these diagnoses to advance criminal justice interventions that align with individuals' political ends. These strategies include using data to reform the police that generated the biased data in the first place, such as by allocating police patrols more equitably; 
reaffirm or strengthen existing models of policing; reject the use of predictive policing tools; 
and reframe the narrative away from reforms like data-driven policing toward abolitionist aims, such as repairing communities.

We identify the implicit assumptions and scope of these varied uses of algorithmic bias as evidence, showing that they require different (and sometimes conflicting) values about policing and AI. Notably, using enforcement bias to inform more efficient allocation of police patrols diverges from the use and interpretation of the same evidence by stakeholders with lived experience in the criminal legal system, who diagnose the source of the bias as systemic. This divergence reflects long-standing tensions in the criminal justice reform landscape between the abolitionist values of liberation and healing often centered by system-impacted communities and the values of surveillance and deterrence often instantiated in data-driven reform measures. We discuss the implications of these differences in relation to participants' social positions, and we advocate for centering the interests and experiential knowledge of communities impacted by incarceration through participatory methods to ensure that algorithmic bias evidence can function as a device to challenge the status quo. 


\section{Background and Related Work}

\subsection{Algorithmic bias: from an object of social critique to a device for social change}

Notwithstanding the widespread claim that algorithmic bias is a mirror of existing disparities \cite{mayson_bias_nodate}, it is rarer for algorithmic bias to be taken seriously as a source of evidence to emancipate and support marginalized communities. Bias has helpfully been identified as a boundary object \cite{grill_bias_2023}, an entity which, despite disagreement about its nature, provides a lens for critique and confrontation of different stakeholders' views.  In this paper, we take this characterization of bias a step further; namely,  we want to move from bias as an object of and for critique to a pragmatic device to address the disparities it mirrors. 

Skepticism about applying a computational lens to societal problems \cite{broussard_artificial_2019, dignazio_counting_2024, eubanks_automating_2018, gebru_oxford_2019, green_data_2021, benjamin_race_2019} has led to a reconsideration of the roles of computational tools from optimization and individual predictions toward systemic interventions to address deeper issues of justice and equity \cite{barabas_interventions_2018, green_data_2021, barabas_refusal_2022}. Recent work in critical data studies \cite{abebe_roles_2020, dignazio_counting_2024, passi2019problem}, as well as past work at the intersection of sociology and statistics \cite{bruno_statactivism_2014, desrosieres1998politics} has mapped out ways in which practices around the use of computational tools, from the design to the deployment of these tools, can be used to address existing disparities or injustices in the status quo.

Computational tools can help unveil or criticize specific aspects of society. Through their reliance on historical and empirical data, tools both in statistics \cite{bruno_statactivism_2014} and in machine learning \cite{abebe_roles_2020} can help disclose the wrongdoings of an institution by appropriating or questioning its own numbers. The categorization and labeling needed to make data legible can indirectly lead to the creation or resurfacing of neglected or unrecognized social categories,  
redefine an existing category, or provide a means to defend a category and its rights (e.g., make it ``legible'' to the system) \cite{bruno_statactivism_2014, dignazio_counting_2024}. 
The use of computational tools can also reformulate social problems and reshape public objectives. The exercise of problem formulation and formalization the tools entail can help reason about and reassign an institution's priorities \cite{bruno_statactivism_2014}, articulate and reason about how societal problems are formulated \cite{passi2019problem},  and highlight the limitations presented by computational techniques \cite{abebe_roles_2020}. 

The power of numerical tools, however, can work in both directions. It can expose as much as it can affirm existing power structures \cite{bruno_statactivism_2014, barabas_refusal_2022}. The same numbers that can be used by communities to expose their oppression can, for example, be used by police to justify their intervention. 
Starting from these considerations, this article explores whether algorithmic bias can be used as a device to disclose or address the societal disparities that it mirrors. At the same time, we explore how evidence of bias enables diverse problem diagnoses and supports the identification of solutions contingent upon individual worldviews. Given its potential to disclose as much as affirm power, we explore how -- if at all -- this evidence can serve the interests of affected communities.

We connect these questions to a parallel tension in criminal justice between abolitionist and reformist responses to the crisis of mass incarceration. Abolitionist philosophy calls for an end to the reliance on imprisonment and policing and proposes alternatives that engage holistically with the structural conditions underlying violence and suffering 
\cite{mcleod2018envisioning}. In this vein, abolitionists often advocate for `non-reformist reform',  change that is not cordoned by what is possible within the current system but rather change that is premised on what should be made possible long-term given human needs 
\cite{gorz1968strategy}. Non-reformist reform prioritizes decarceration and reinvestment of resources back into communities. This is in contrast to `reformist reform', which limits its objectives to the maintenance and practicality of the current system. Most data science applications in the context of criminal justice are reformist reforms \cite{green2019good}; as we see in the interview results, the predictive policing tool itself is seen as a reformist reform, drawing a sharp line with reform advocates in favor and abolitionists opposed. 


\subsection{The Chicago Crime Prediction Tool}

Since the early 2000s, police departments around the US have begun adopting new technologies to forecast crime, often in partnership with private companies like PredPol and Palantir \cite{brayne_predict_2020}. The Chicago Police Department (CPD) first began using CompStat, an early data-driven geographic forecasting policing program, in 2003 \cite{verma_heat_2021}. In 2012, the CPD started using the Strategic Subject List, a person-based predictive policing tool made by the Illinois Institute of Technology, to generate a ``heat list'' of likely perpetrators or victims of violent crime  \cite{stroud_automated_2021}. The program was terminated in 2019 following investigations showing that the program was inaccurate, ineffective, and racially biased, among other problems \cite{kunichoff_contradictions_2017, sweeney_for_2020}. In 2018, the CPD started using ShotSpotter, a controversial gunshot audio detection technology, to dispatch police to areas with suspected recent gunfire, which led to high-intensity investigatory stops and fatal interactions between the CPD and city residents \cite{simkin_looking_2023}. In March 2021, ShotSpotter was involved in an incident in which police killed 13-year-old Adam Toledo, intensifying community opposition to the technology. After investigations found that Shotspotter was inaccurate, exacerbated the over-policing of neighborhoods of color, and increased unconstitutional stop and frisk by police \cite{ferguson2021chicago, feathers_gunshot-detecting_2021}, city officials announced that they would end their contract with the company by September 2024 \cite{mohtasham_chicago_2024}. 
These events all happened in the backdrop of the 2017 consent decree, a court order mandating broad police reform in the city of Chicago.  



A recent article on event-level prediction of urban crime in Chicago (herein referred to as the Chicago crime prediction algorithm) \cite{rotaru_event-level_2022} is the latest development in this spotted history, though the algorithm is not yet deployed by any police department. Using violent and property crime incidence data from the city of Chicago, Rotaru et al.\ built a spatiotemporal network that inferred patterns from past event occurrences. This network, known as the Granger network, comprised local estimators forming a communicating network to predict future infractions. Using publicly recorded historical event logs, Rotaru et al.\ extended their analysis beyond the training sample to events in the subsequent year. See \autoref{figure:model_card} for a model card \cite{mitchell_model_2019} outlining additional details of the algorithm.  

In contrast to the CPD's wave of failed predictive policing experiments, Rotaru et al. present their novel algorithm as taking a different approach to predictive policing. Their stated aim is not simply to predict hotspots of crime with greater accuracy than commercially available tools but also to ``garner deep insight into the nature of the dynamical processes through which policing and crime co-evolve in urban spaces,'' including how policing interacts with, modulates and reinforces crime. They emphasize that 
they see their algorithm as serving the dual role of enhancing state power through criminal surveillance while also providing a means to scrutinize states by allowing researchers to ``audit them for enforcement biases'' \cite{rotaru_event-level_2022}. Indeed, 
Rotaru et al.\ write that the Chicago crime prediction algorithm indicates the presence of socioeconomic bias in law enforcement. Modeling the number of individuals arrested during each recorded event allowed the researchers to 
contrast crime rates with arrest rates, revealing that increases in crime rates correlated with increased arrests in wealthier areas, whereas in low socioeconomic status 
neighborhoods, arrest rates dropped without a corresponding decrease in crime. This pattern persisted across all the years under analysis. The researchers attributed this bias to resource constraints on police, compounded by discretionary prioritization of wealthier neighborhoods, corroborating suspected enforcement bias toward US suburbs with high socioeconomic status \cite{meyer_suburban_2016, lipton_why_1977}. The authors used this evidence of socioeconomic enforcement bias to suggest a need-based reallocation of police patrols across neighborhoods in Chicago.


\begin{figure}
\caption[Model Card]{Chicago Crime Prediction Algorithm Model Card}
    \fbox{\parbox{8.3cm}{
 \noindent \begin{itemize}[leftmargin=*]
\small 
    \item \textbf{Model Details}
    \begin{itemize}
        \item Spatio-temporal network inference algorithm: infers patterns of past event occurrences and constructs a communicating network (Granger network) of local estimators to predict future infractions.
        \item Published by academic researchers Rotaru et al.\ at the University of Chicago in \textit{Nature Human Behavior} in 2022 \cite{rotaru_event-level_2022}.
    \end{itemize}
    \item \textbf{Intended Use}
    \begin{itemize}
        \item To accurately predict hotspots of crime in Chicago, producing insights on the processes of crime and its interaction with policing. 
        \item The authors recommend policy changes for ``more equitable, need-based resource allocation'' of police and suggest that their tool can ``be used to track the extent to which such policies approach this trace of equitable enforcement allocation.''
    \end{itemize}
    \item \textbf{Training Data}
        \begin{itemize}
            \item Spatiotemporal logs of crime incidence data in Chicago, as well as the nature, category and description of each incident.
            \begin{itemize}
                \item Public domain criminal event logs were also analyzed for seven additional cities: Detroit, Philadelphia, Atlanta, Austin, San Francisco, Los Angeles and Portland.
            \end{itemize}
            \item Socioeconomic variables corresponding to regions of Chicago.
            \item The number of individuals arrested during each recorded event is modelled separately, allowing the investigation of enforcement bias.

        \end{itemize}
            
    \item \textbf{Metrics and Evaluation}
        \begin{itemize}
            \item Area Under Curve (AUC)
            , Predictive Accuracy Index (PAI) (normalized event rate in identified hotspots) 
            , Prediction Efficiency Index (PEI). 
            \item  PAI and PEI results were compared against the best-performing teams in a crime forecast challenge hosted by the National Institute of Justice in 2017. 
            \item Enforcement bias: ``In wealthier neighbourhoods, the response to elevated crime rates is increased arrests, while arrest rates in disadvantaged neighbourhoods drop but the converse does not occur.''

        \end{itemize} 
        \item \textbf{Ethical Considerations}
        \begin{itemize}
            \item Potential misuse: the authors acknowledge that these results may be falsely interpreted to justify increased police intervention in higher-crime (often more racially and ethnically diverse) neighborhoods, which could lead to increased surveillance and fatal police interactions in marginalized communitiees.
            \item Bias in crime reporting data: Rotaru et al.\ also note that incident logs likely contain ``biases arising from disproportionate crime reportage and surveillance [of communities of color],'' which ``no amount of scrubbing or clever statistical controls can reliably erase.'' 
        \end{itemize}
        
            

\end{itemize}}}
\label{figure:model_card}
\end{figure}

\vspace{-.4em}
\section{Methods}

To interrogate the ways in which the enforcement bias captured by the Chicago crime prediction algorithm can serve as evidence, we conducted semi-structured interviews with 18 stakeholders, primarily from the Chicago metropolitan area. 

\noindent \textbf{Recruitment and Demographics.} We conducted interviews with stakeholders from various backgrounds, including academia, community organizations, and the public sector. After an initial mapping of relevant actors through desk research, we selected the rest through referrals from initial respondents. At the time of this writing, the Chicago crime prediction algorithm has not been publicly deployed; snowball sampling allowed us to trace a latent network of relevant actors that we otherwise would not have identified \cite{parker_snowball_2019}. We continued recruiting and interviewing participants from each background until no new themes or information directly relevant to the project emerged from the interviews \cite{small_how_2009} or our timeframe for data collection passed. In total, we attempted to recruit 62 participants, achieving a response rate of 29\%. 
We made an effort to select participants with variation \cite{weiss_learning_1994} across roles (see \autoref{table:demographics}) and level of personal exposure to the criminal legal system (see \autoref{table:cls_exposure}), ranging from no prior involvement to formerly incarcerated. This variation enabled us to consider the role of individuals' positionality in the results; we indicate each quoted participant's background (`C' for community organization, `R' for researcher, or `P' for public sector). While eight participants had direct experience or knowledge of the algorithm, several interviewees had no prior familiarity with it; these latter interviews helped inform our understanding of Chicago-area policing more broadly. 

Due to confidentiality concerns, recruiting participants from the public sector was challenging. This is a common issue in interviews with participants in senior or official roles -- so-called ``elite'' interviews \cite{richards_elite_1996}. We accounted for this by recruiting two additional participants, one who is a contractor for the criminal legal system and another who works for an oversight body. 

\begin{table}[t!]
		\small
		\centering
		
		\begin{tabular}{  l c r  } 
			
			\textbf{Role} & \textbf{Frequency} & \textbf{\%} \\ 
			
			\hline

			\textbf{Community Organizations (C)}&7&38.8\%\\
            \hspace{1.5em}Re-entry Associations&4&22.2\%\\
            \hspace{1.5em}Safety \& Community Initiatives&3&16.6\%\\
            \hline
            \textbf{Researchers (R)}& 6&33.4\%\\
            
            \hspace{1.5em}Tool Developers&1&5.6\%\\
            \hspace{1.5em}Academic Researchers&5&27.8\%\\
            \hline
            \textbf{Public Sector (P)}&5&27.8\%\\
            
            \hspace{1.5em}Public Officials&1&5.6\%\\
            \hspace{1.5em}Police&2&11.1\%\\
            \hspace{1.5em}Contractors \& Oversight Bodies\footnote{This refers to people such as contractors and consultants working with or for criminal legal system institutions and independent bodies working as a ``broker'' or as a link between communities and the criminal legal system.}&2&11.1\%\\
            Total&18&100\%\\
			\hline
		\end{tabular}
		
		\caption[Roles of participant population]{Roles of participant population.}
		
		
		\label{table:demographics}
	\end{table} 

\begin{table}[t!]
		\small
		\centering
		
		\begin{tabular}{  l c r  } 
			
			\textbf{Exposure to the criminal legal system} & \textbf{Frequency} & \textbf{\%} \\ 
			
			\hline
	
			System-impacted individuals&2&11.1\%\\

            Working directly with system-impacted individuals&7&38.9\%\\

            Employed by the criminal legal system (e.g., police)&2&11.1\%\\

            Working with the criminal legal system&3&16.7\%\\
            
            No direct exposure (e.g., researchers) &4&22.2\%\\
            Total & 18 & 100\%\\
            \hline
            
		\end{tabular}
		
		\caption[Participant exposure to the criminal legal system]{Participant exposure to the criminal legal system}
		
		
		\label{table:cls_exposure}
	\end{table} 

\noindent \textbf{Interview Process.} Interviews ranged from 30 minutes to 1 hour, with an average length of 39 minutes. The first author conducted all interviews; two interviews were carried out in person and the rest via video call. Interviews were semi-structured  \cite{edwards_what_2013}, allowing us to focus and dig deeply into an initial set of topics, while also letting participants' expertise guide the interview and allowing new themes to emerge \cite{roulston_sage_2018}. Specifically, we asked about participants' understanding of and attitudes toward algorithmic bias generally and police enforcement bias specifically; the challenges and problems they believe to be associated with these biases; and the interventions available or those they envision in the future (see \autoref{sec:interview_guide} for a full list of interview questions). We adjusted the content and wording of the questions slightly depending on the participants' particular knowledge of or experience with the algorithm and the criminal legal system. For example, for participants with intimate knowledge of the algorithm, such as its developers, more technically detailed questions were posed, while for participants with local knowledge of the criminal legal system, such as members of community organizations, a stronger emphasis was placed on questions about the Chicago context and community safety.

\noindent \textbf{Qualitative Analysis.} We analyzed interview transcripts using the qualitative data analysis software Nvivo. Both authors analyzed interviews iteratively, identifying recurring themes, reviewing transcriptions 2-3 times, and switching between inductive coding and data collection to clarify and strengthen themes and uncover opposing evidence \cite{miles_qualitative_2014}. We converged on seven high-level themes (in order of most to least prevalence in the interview data): downsides of AI; upsides of AI; structural bias; lived experience and positionality; abolition; reformism; and controversy around policing (see \autoref{sec:code_table} for a code table). We complemented the results from the interviews with a document analysis \cite{karppinen_what_2011} of policy proposals and programs produced by the interviewed stakeholders, as well as other active participants in the domain of policing and justice reform in the Chicago area (e.g., policymakers, regulators, community organizations, and think tanks). Through triangulation with findings from our interview data, we were able to elucidate, refute, or expand on the findings from our interview data \cite{frey_sage_2018}. 

\section{Results}




Across the board, interviewed stakeholders strategically used algorithmic bias as evidence to advance interventions that aligned with their own positionality and values about the criminal legal system. Drawing inspiration from Catherine D'Ignazio's taxonomy of feminicide activists ``refusing and using'' data \cite{dignazio_counting_2024}, we see that stakeholders from different backgrounds use evidence of algorithmic bias to \textbf{reform} the policies around police patrol allocation; in the case of more powerful actors, \textbf{reaffirm} their own authority or existing power structures; \textbf{reject} algorithm-based policing interventions; \textbf{reveal} counterdata on authority figures in an effort to subvert their power; \textbf{reframe} crime as a structural rather than interpersonal problem; and \textbf{repair} and heal families and communities. 



\subsection{Reform: Towards ``a legitimate, ethical, fair police force.'' [P013]} 
\label{sec:reform}

\subsubsection{Policing as an optimization problem.}
\label{sec:allocation} Making policing more `fair' is a standard goal of criminal justice reform \cite{van2017achieving}. Stakeholders from academic and public sector contexts tended to focus on `over-' or `under-policing' as the source of the Chicago crime prediction algorithm's bias, 
emphasizing the benefits of allocating policing more optimally, such as increased public safety and police legitimacy. However, these participants had mixed views about whether predictive policing was a viable strategy to achieve this goal. 

Three of the academic researchers we interviewed, including one of the tool's developers and members of a Chicago-area research group focused on criminal justice research, framed the problem revealed by algorithmic bias as a police allocation issue. One senior member of the research group [R006] explained the need for improving police presence in marginalized neighborhoods: 
``People who live in those communities ask for more police because they're facing high levels [of violent crime] ... 
I mean, the reality is like somehow we are both, like, those communities are both over- and under-policed at the same time.'' Another member of the research group [R007] posited that ``police presence alone deters crime,'' even if police make no arrests. 

These attitudes aligned with perspectives voiced by two police officers. Echoing the senior academic's perspective, one officer [P003] argued that ``the residents in those [Black and Brown] neighborhoods, that's the silent majority. They want us there, they need us.'' Another officer [P013] distinguished ``feeling safe'' and ``being safe,'' emphasizing the significance of the former and highlighting that police presence contributes to a sense of security among citizens. 
On this shared view, there is intrinsic value to policing, and distributing it optimally is a primary reform goal. 

However, some stakeholders who were otherwise proponents of reforming police allocation expressed reservations about the use of predictive policing specifically to achieve those goals. The officer who argued that increased police presence promotes the feeling of safety [P013] warned that the use of predictive policing and other technological systems could be ``preventing cops from doing their job, which is a social job'' by making them too ``comfortable coming into an office and sitting behind a computer all day,'' thereby diminishing the physical presence of police in communities and compromising the feeling of safety. Likewise, one of the tool's developers [R002] did not want the tool to be directly used to inform police allocation decisions.

\subsubsection{Reducing interpersonal bias.} \label{sec:412} Several public sector actors saw the tool's enforcement bias as a potential result of another kind of unfairness -- interpersonal bias, including implicit bias, the bad judgment of a ``few bad apples,'' or simply limitations in human cognition. Reducing interpersonal bias among police officers through implicit bias training, data-driven policing, or more diverse police forces is another standard route of reform \cite{spencer2016implicit, van2017achieving}.  

One county commissioner [P015] in a majority Latino district of Chicago called implicit bias ``very damaging'' to communities and explained that police officers, judges, and ``all of our criminal justice system'' must attend implicit bias training. 
A police officer [P013] also described what he called declining ``standards of professional behavior'' among police officers, underscoring the need for strong leadership programs, adding that this would improve community trust and legitimacy of police. 

Arguing for the benefits of ``research-driven'' policing, another police officer [P003] admitted that ``we're not the brightest bulbs'' but that ``there are smarter minds analyzing all the different mathematical models in the back end of the AI, so they'll be able to inform us and they'll keep all of the AI models accountable.'' Using the example of people's tendency to overcount the significance of a stale criminal record, one academic researcher [R006] argued that an algorithm could ``have a sort of more built-in decay mechanism than like a human could,'' thus improving on limitations in human cognition.

 

Participants across all roles also proposed diversity, equity and inclusion (DEI) as a strategy for minimizing interpersonal bias, especially integrating diverse perspective or other stakeholders in the creation of data and algorithms.  
As one interviewee from a reentry association [C001] put it, ``whose voice should be in the room for the algorithms that are being built, right? And I don't just mean a little bit of, yeah, engaging in participatory research where you sit down with a group of folks and this and that and you go off in your room and you build this thing. No, I mean you actually have to make them or the folks from these communities where these tools are going to be deployed an active creator of the system, right?'' [C001] Several researchers working with system-impacted individuals and victims of violence also stressed the importance of incorporating those groups' voices in research. 

Another participant from a reentry association [C004], however, highlighted that incorporating diverse stakeholders on its own would not solve the problem: ``You could take a Black person or a Brown person, or some other disenfranchised individual, and they could be the author of, you know, some new algorithm. But unless they have a specific training and knowledge around systems of bias -- how they're constructed, how they operate and how they can be deconstructed -- then I don't think that these individuals that are constructing these programs can, can effectively create solutions for them.'' 




\subsubsection{Increasing transparency and accountability.}  \label{sec:transparency}

Beyond algorithmic bias, some participants associated another reform goal with the predictive policing tool, namely, accountability and transparency. 
This is a common byline in police reform rhetoric \cite{white2021can} and is often presented as a justification for data-driven policing and surveillance through body-worn cameras.

Referring to the Chicago crime prediction tool, one police officer [P003] said that ``it being open source is a winner'' because of what he pejoratively referred to as the ``two big buzzwords'' around policing following George Floyd's murder in 2020, namely, accountability and transparency. 
``Here's the open source data, you know. 
You look at it then I'm not hiding anything. There's no secret sauce to this. There's no ulterior motives.''  
%

However, other participants pointed out that the relationship between open-source and accountability is far from a given. One reentry community organizer [C001] argued that ``the people that build them [algorithms] should have some accountability for how they're used ... even if you made it open source, right? Even more then, in some ways, right? You have an obligation to have thought about how it could be used for better or for worse, right?''  One of the tool's developers [R002] acknowledged the same issue with open-source, noting that ``you have to also be careful when you're giving it away''  
because nothing prevents private companies from forking open-source software into a private (i.e., non-transparent) repository. The fact that police officers explicitly reported wanting to acquire the predictive policing tool underscores that even open-source software, once developed, may be used in ways contrary to the desires of their developers, particularly once it becomes commercialized \cite{widder2022limits, widder2023dislocated}.

\subsection{Reaffirm: ``How can there be a bias if in that triangle it's all the same race?'' [P003]}
\label{sec:reaffirm}

\subsubsection{Following the data.} Participants in public sector positions denied the presence of bias or used evidence of algorithmic bias to reaffirm their own authority and the legitimacy of existing policing practices. This is exemplified well by one officer [P003], who candidly explained why there is no over-policing problem: ``Where the crime is happening, the concentration is 99\% Black or Brown and the victims are 99\% Black or Brown, so that's our baseline. ... 
We're there because the data says that's where we need to be.'' He added that ``
it's not that the AI is biased. The AI is just working with the data, and the data says that this geography, this demographic is, you know, these individuals.'' 

This officer provided a problem diagnosis of the algorithm's behavior that essentialized the policed community -- ``these individuals'' -- as inherently exhibiting more criminal behavior, thereby using the pattern to justify or take as given the status quo. Disparities in crime, in other words, are the factor that drives disparities in policing and in turn the algorithmic non-bias, and the officer lamented that ``in policing we've been knocked around'' for simply responding to what ``the data is telling us.'' 


\subsubsection{New and improved technology.}
\label{sec:422}
Another response to algorithmic bias that reaffirms police power is the call to invest in more technology. Some public sector participants saw the algorithm as an opportunity to do their work better, and others as the only way forward. For instance, speaking about the motivation behind the development of the crime prediction tool, one developer explained that ``all of them [previous predictive policing algorithms] were doing it wrong'' and the need to ``show what is the right way to do it.'' 
One police officer [P003] 
called for the need for policing to be informed by the crime prediction algorithm and to ``try a different approach on the AI.''

\subsubsection{More policing.} As we discuss in \S\ref{sec:allocation}, one response to enforcement bias is the call to increase police presence in areas with higher crime incidence, which implicitly affirms the legitimacy of the current policing system. For instance, one police officer reported being interested in ``using this technology to break down the cycles that have been perpetuated for years'' [P003]. However, it became clear that what they meant by this was sending more police to Black and Brown neighbourhoods in order to deter crime, rather than to change the structure of the current system or ``break down'' any cycles. 

An interviewee from a community association [C011] voiced concern about affected communities similarly endorsing approaches that do not change the existing system and go against their own interests, such as ``more policing,'' and attributed this to a lack of awareness of alternative options. Referring to the race for a major election in a large US city, they said, ``... older Black residents who are concerned about public safety, and many of them, their response is `we need more police'. And the reason is because it's the only tool they know, right?''  


\subsection{Reject: ``I don't think it should be used in any way, shape or form.'' [C005]} 

\subsubsection{Garbage in, garbage out.}
\label{sec:garbage}
Stakeholders across all groups shared some inherent reservations about AI in the policing context, particularly concerns about its underlying data. In particular, the theme of algorithmic bias being a ``result of the system'' [C001] -- that is, reflecting structural bias and systemic racism -- recurred frequently. 
One system-impacted participant [C005] argued that the system should not be used at all because it does not account for systemic and individual bias: 
``It just cannot be neutral, and therefore it should not be used because people that look like me are the ones who are going to be impacted.'' 

Another concern voiced by researchers working with system-impacted individuals was a general distrust of low-quality or manipulated police data. ``Their [police] incentives are wrapped up in what is reported. You know what crimes are in the news and what people are saying. It's like they're kind of incentivized to obscure data.'' [R008] Researcher participants described that documentation of police misconduct is likely to be reduced to the most minor violation; for instance, an illegal search will only be reported if it also includes sexual assault. 
Another researcher working with system-impacted individuals [R009] pointed to several cases where police officers intentionally misreported the race of individuals in order to conceal bias and discrimination in their practices. He mentioned that in Louisiana, for example, several law enforcement agencies targeted Latino individuals but identified them as white on tickets in order to conceal discrimination \cite{webster2021if}, and similar incidents have also taken place in Connecticut \cite{nierenberg_over_2023}, Los Angeles and San Francisco \cite{barba_are_2023}.  

Finally, several participants from commmunity organizations had reservations about the inherently interpretative component of AI, which they said biases its results. A member from a reentry community organization [C001] argued that AI is an 
``interpretive instrument'' rather than a mechanical one. ``In an ideal sense, let's say the [AI] system had no politics, right? But the people that will use the results and interpret the information do, and you know, as they, you know, they always say, you know, people can make statistics tell any story they want it to tell.''

\subsubsection{Unneeded, unconvincing, and resource-intensive.}
\label{sec:unneeded}
Other respondents were opposed to the predictive policing tool for more pragmatic reasons. For instance, several participants felt that the algorithm's enforcement bias did not show anything that was not already known. One academic researcher [R006] commented  that developing an algorithm is ``a pretty inefficient way'' to inform police reallocation, arguing that ``
you wouldn't need that, to develop an algorithm to reach these same conclusions.''  An interviewee working for a reentry association [C017] expressed similar skepticism about the need for a tool that predicts crime. ``He [system-impacted individual] is always going to be discriminated against in one form or fashion, maybe an employment house. He's always going to be there, and that's a predictor of recidivism or returning to be a criminal. Well, yeah. We don't need an algorithm to say that.'' 

The academic researcher [R006] also said that the algorithm was ``not particularly convincing'' for other stakeholders: ``I barely understand exactly what they're doing and how they reach their conclusion, ... And like, I have a PhD and I do this for a living. So I think, you know, like trying to convince like, other stakeholders is very difficult.'' Other participants, including another academic researcher and several community organizations, also reported not understanding the paper or how the tool worked, though in the case of one police officer [P003], that was a positive feature because it indicated that the tool's developers were ``smarter minds.'' 

The same police officer [P003] also commented on the resource-intensiveness, both in terms of computational power and departmental funding, of acquiring and using a predictive policing tool in practice. 
%
``Will we have a budget for this? Or do I need to go out and fundraise for it and convince foundations that, you know, this is the way we need to use the limited resources that we have?'' 

Understanding the underlying motivation for developing the tool helps contextualize the above criticisms. One of the tool's developers [R002] explained that the team had been working on a general algorithm and were looking for a real-world test case to demonstrate how the algorithm could be instantiated. ``Why crime? No particular reason. We were looking for a problem that would be an impactful thing.'' An advantage of working on crime, they add, is that there is ``lots of data on it.'' The research team had access to compute power that a typical community organization or police force does not, and it is clear that the team did not have the goal of convincing stakeholders or revealing surprising patterns. Rather, the research team saw predicting crime as an ``interesting'' challenge, seeking the `technical sweetness' \cite{hill2023your} that engineers feel when they encounter the solution to a challenging technical puzzle. 

\subsection{Reveal: ``Data that helps us understand how the police see our communities.'' [R009]}
\label{sec:reveal}

Enforcement bias was one type of evidence community organizations used to engage in sousveillance, or `watching from below', in which the traditional framing of panoptic state surveillance is inverted, allowing citizens to observe powerful state actors \cite{mann2003sousveillance, newell_introduction_2020}. Researchers and community organizers described how they co-opted surveillance methods to watch the watchers. Alternate data sources, or counterdata \cite{dignazio_counting_2024}, reveal issues with policing, remediate problems with how police and victim data are collected and reported, and allow alternate narratives to emerge. 

One member of a community safety initiative [C101]  lamented that 
``a lot of times what criminologists will tend to do is they'll just look at the data. And I think what's missing is that there's a story behind it.'' To provide this missing context, one human rights researcher [R009] described working with grassroots groups that do on-the-ground documentation through cop watching and court watching, gathering counterdata on police officers, judges, and prosecutors. They argue that ``you need to record different things than what the courts are'' in order to include context normally absent in court data. The same researcher also described using data about human rights violations to understand patterns of violence, and even to make statistical estimates of events that are not represented in the data. 



As we discuss in \S\ref{sec:garbage}, several researchers working with system-impacted individuals also reported the existence of low-quality, unreliable or manipulated police data. 
By collecting data from the Chicago Police Department, the city of Chicago, and the Civilian Office of Police Accountability through Freedom of Information requests, one research group [R008] described how they published an alternative dataset about police misconduct in Chicago. They also recounted collaborating with system-impacted individuals and victims of police violence to build a basic classifier to predict police misconduct using complaint data generated by community members. This alternate algorithm revealed a culture of abuse within policing: ``
instead of something like operation and personnel violation, you know, we're able to see clusters of, you know, families who are like ... `this is how police treated me'. And what was really fascinating about that project is that we were able to, you know, move away from this idea of like, individual officers, like, back up to what is maybe the culture of policing that exists, what are the kind of tactics that, that people are reporting that connect them across space and time?''


\subsection{Reframe and Repair: ``Deconstructing the status quo and recreating a new status quo.'' [C004]} 
\label{sec:reframe}

Countering mainstream understandings of crime among respondents from the public sector and some academic researchers, many participants from community organizations presented alternate data epistemologies \cite{ricaurte2019data} focused on abolitionist understandings of the ``root of the problem'' [C016]. 
In this vein, participants responded to enforcement bias by calling for the need to recognize structural racism, reframe the narrative toward an understanding of crime as structural rather than interpersonal, and repair and build healthy communities.


\subsubsection{Recognition of systemic racism.}
\label{sec:recognition}

A common thread in the interview data was the ``tale of two cities,'' [C001; P015] or Chicago's history of segregation and systematic disinvestment in its Black and Brown communities. Participants emphasized the significance of Chicago's history of segregation, citing it as a root cause of not only crime but also numerous other social problems, such as disparities in life expectancy -- ``I want to say 20-30 year difference in life span between those that live in the poor neighbour community'' [C001] -- and in wealth --``You can go to the different parts of the north side and you can just see the wealth, right? ... and then you can go to the south side or the southwest side and see completely different-- you, you can almost think that you're in a different city'' [P015]. 

Some system-impacted individuals and researchers working with them also remarked on the intrinsic suspicion they experience toward policed communities, accompanied by a tendency to trust or believe those doing the policing. One of them [C017] argued that criminal legal policies are ``setting people up for failure,'' using Nixon's 1970s policy on potential delinquents as an example of how `criminals' are created out of assumptions rather than facts.

Multiple participants stressed that the first step to resolving the problem presented by enforcement bias is having a clear understanding of what the problem is. One member of a community safety initiative in Chicago [C011] noted that ``before anybody can heal, there needs to be a reckoning, there needs to be some kind of identification that, that the systems that were created actually have caused harm.'' Two more participants [C004; C005] echoed this sentiment almost verbatim, arguing that ``if we're not going to address a system of white supremacy, then we're not going to get to any real hard solutions.'' [C004] 

\subsubsection{Narrative change.}

Beyond reckoning, participants emphasized reframing the narrative around system-impacted individuals, police, and community safety.  As one community worker [C005] put it, ``The delusion that somehow we can scare people out of crimes of poverty is ridiculous. We have been so indoctrinated by current systems to accept the claims, regardless of the the reality, right?'' They argued that ``One of the biggest things that we do is narrative change work'' in order to push back on dominant narratives around crime and policing. 

In particular, participants call for a reorientation toward a structural understanding of the ``root cause'' [C014] of crime, focusing on treating the cause rather than the symptom. Two participants criticized interventions that focus on the interpersonal level, with one participant [C011] comparing the current criminal legal system to a hospital that only contains an emergency room and another [R104] lamenting that interventions like community violence and prevention programs ``stop at the interpersonal level, but they don't also account for the structural systemic impediments at play.'' 

\subsubsection{``Building healthy communities rather than reducing crime'' [C001]}


Grounded in an  alternate understanding of crime as structural, participants responded to enforcement bias by advocating for non-reformist reforms such as alternatives to police and rethinking how to use data to promote community safety. 

One community organizer [C001] asked, ``instead of thinking about how we reduce violence and crime, what if we pose the question of how this information helps us build healthy communities?'' We saw in \S\ref{sec:reveal} how researchers and activists gather counterdata on police and make up for limitations in data. This strategy centers the needs of affected communities, which community organization participants agreed got little benefit from predicting crime.  Another researcher [R008] explained how AI could instead be used to reduce crime in the sense of supporting survivors of violence, especially police violence: ``When we pivot our approach to, you know, engaging with survivors, then we're better understanding the implications of policing and the ways that we can use artificial intelligence to support survivors of violence and reduce crime, like, period.'' 


Beyond the use of data, several participants also advocated for non-reformist reforms. Drawing on an abolitionist critique of police, one researcher [R008] working with system-impacted individuals challenged the assumptions underlying policing:
``Policing should not be about `we gotta get the bad guys', right? It should be about `how do we keep people safe'.'' Taking this view a step further, one community organizer [C101] questioned the belief that ``we need more [police],'' adding that ``if we bring more tools to the community, there's-- the solutions are here. We just haven't given you all the equipment to do so.'' Notably, the police and academic position on police promoting safety was starkly different from the perspective of some interviewees from community organizations for which police presence was an ``enforcement presence, not a-- not a deterrent, but an enforcement'' [C001].



Even when there was agreement about the need to strengthen communities, however, respondents differed in their views on which kinds of interventions to put in place to do so. Some reentry organizations advocated for giving the same rights to system-impacted individuals or clearing their past criminal record after some time [C017], while others fought for more police and state accountability [R008]. One individual working as a contractor for the public sector [P016] recognized the structural nature of the problem and yet focused on ``changing the individual'' as a solution, namely, moving individuals out of a high-crime environment or, alternatively, strategically reclaiming high-crime spaces by organizing community events (e.g., sports competitions, community barbecues, etc.). Another member of a community organization [C011] focused on ``changing the conditions'' -- namely, focusing interventions on the socioeconomic determinants of crime such as wealth and health. In short, even within the non-reformist reform branch of responses to enforcement bias, there were a diversity of perspectives on the appropriate kinds of interventions.

\section{Discussion}

Our results shed light on the socially contested nature of algorithmic bias within the realm of predictive policing, illustrating that bias is not understood as a singular challenge but rather evidence of a multifaceted package of problems with a corresponding broad range of possible interventions. We grouped the interview findings according to whether stakeholders from different backgrounds used evidence of algorithmic bias to reform, reaffirm, reject, reveal, reframe or repair, showing the ability of enforcement bias to be used both to support and challenge the status quo. 

Whether and how algorithmic bias was seen as useful for advancing each of these goals varied according to each stakeholder's broader understanding of the source of police enforcement disparities -- structural inequality versus interpersonal bias -- and the set of interventions aligned with each respective problem diagnosis. Stakeholders from different professional roles and positionalities tended to articulate problem diagnoses of the bias aligned with their own relationship to the criminal legal system, though we saw a diversity of attitudes and responses within each subgroup and points of convergence between individuals with different positionalities and roles. For instance, we saw substantial overlap among interviewees from all backgrounds regarding the importance of community safety and potential downsides to the use of AI and data in the context of policing, such as AI's possibility to diminish the role of nuanced human discretion. On the other hand, we saw disagreement between groups about whether police presence promotes community safety, and there were differences within groups relating to whether predictive policing tools would improve policing and whether incorporating affected stakeholders in the development of predictive policing  could improve the tools' viability. 



Many of the inter-group differences we saw reflect tensions between the assumptions underlying interventions grounded in reform versus abolition -- reformist reform that operates within the logic of the current system and non-reformist reform that aims for longer-term systemic change, respectively. Along these lines, we saw other contested concepts emerge, such as the notion of community safety and `overpoliced' communities, with the connotation and valence of `overpoliced' changing depending on the socioeconomic status of the overpoliced community. For example, in their paper, developers of the Chicago crime prediction algorithm suggest that  police are significantly more present in wealthier neighborhoods because those areas have more resources. 
`Overpoliced' here has a positive connotation -- the privilege and safety of well-resourced communities. At the same time, interviewees from community organizations and from the public sector tended to refer to overpoliced communities as poorer communities with higher rates of crime. 

While inter-group differences reflect historical tensions between reform and abolition, one additional significant binary emerged that correlated with participants' attitudes toward the criminal legal system -- `AI hype' rhetoric. We examine this finding below and we end by discussing the relationship between participant positionality and research on algorithmic bias in predictive policing. 

\subsection{AI hype rhetoric}

With stakeholders from all groups, except the academic researchers most closely familiar with the tool's development, attitudes toward predictive policing tended to fall into one of two AI hype narratives: optimism about AI's potential to `disrupt' the criminal legal system in a beneficial way, and inherent opposition to AI on the basis that it is racist and fundamentally flawed. Techno-optimism mapped onto the extent to which a stakeholder understood the problems the algorithm was capturing as primarily interpersonal rather than structural, which correlated with positionalities in higher positions of power. However, some endorsers of structural inequality narratives still saw value in certain applications of AI, such as grassroots classifiers to counter narratives put forward by police or to reveal patterns of police violence. 

The clearest example of disruption rhetoric came from officer P003, who praised the 
``smarter minds'' developing the predictive policing tool (\S\ref{sec:412}) and argued that more technology -- or trying ``a different approach on the AI'' [P003] -- could help tackle the bias in the algorithm  (\S\ref{sec:422}). They also  
remarked repeatedly that AI could help the police force's ``marketing ability to brand us as open as transparent and as accountable,'' what they called the two salient ``buzzwords'' of the ``2020, BLM, post-Floyd'' era (\S\ref{sec:transparency}).

AI may yet become a successful marketing tool for police forces, similarly to past cases in which police acquisitions of technologies like body-worn cameras strategically co-opted community calls for accountability to justify larger police budgets \cite{umansky_failed_2023, karakatsanis2019usual}.\footnote{Police are trained to manipulate and withhold body-worn camera footage to their own benefit \cite{umansky_failed_2023}.} However, the underlying claim that predictive policing promotes transparency and accountability in policing in based in fiction. Sociologist Sarah Brayne provides a compelling argument that predictive policing simply displaces discretion to other, less publicly visible stages of the policing process, thereby reducing both transparency and accountability \cite{brayne_predict_2020}. 

On the other end of the AI hype spectrum, we saw community organizers describe AI as inherently racist. As we discussed in \S\ref{sec:garbage},  
a system-impacted leader of a reentry organization [C005] argued that predictive algorithms should not be used at all in the context of the criminal legal system because ``it's just racist and it needs to be stopped.'' Those in positions of power, they argued, are incentivized to use algorithmic systems because ``it keeps the status quo in place,'' which is what gives those individuals power. An interviewee running a reentry association [C017] echoed that these systems ``set people up for failure'' (\S\ref{sec:recognition}) and that people with a criminal record are ``always going to be discriminated against in one form or fashion'' (\S\ref{sec:unneeded}). 
Like P003, C005's and C017's perspective paints a broad and uniform brush across all uses of AI in the context of the criminal legal system. Both groups at the extreme ends of the spectrum had direct experience with the police but from starkly different positions of power; from being a police officer to having been previously incarcerated.

Not all the participants we interviewed fell into this AI hype dichotomy, but the polarity was widespread enough that we believe it is important to explicitly take into account in future work that aims to use algorithmic bias evidence to challenge the status quo. In practice, existing empirical research of evidence-based reforms shows that, despite their hype and high price tag, they often have minimal effects that wash out over time, suggesting that these reforms are indeed harmful but not in the ways that critics may expect \cite{pruss2023ghosting, stevenson_assessing_2018}. As we explain in the following section, participatory research is crucial to develop responses to algorithmic bias that are attentive to the interests and knowledge of communities impacted by incarceration; at the same time, such work would be made more robust by helping participants resist one-dimensional narratives about the harms of technology in the criminal legal system. 

\subsection{``Bias as defined by who, right? So ultimately, who's the judge?'' [C001]}

The range of problem diagnoses and uses of algorithmic bias might be interpreted as a `relativistic' stance toward the issue, where the understanding of algorithmic bias is contingent upon an individual's subjective world view. Rather than dismissing these differences, we believe that they provide a valuable lens to understand the conditions under which algorithmic bias becomes a device to reinforce or challenge power dynamics. Drawing on standpoint theory \cite{harding_rethinking_1992}, we argue that unraveling the relationship between bias and positionality serves as a critical tool to evaluate shortcomings and mistaken assumptions in potential responses to algorithmic bias coming from actors in relatively high positions of power, such as police officers and even academic researchers such as ourselves. 

Standpoint theory posits that individuals' social positions shape their access to knowledge, influencing how they interpret and respond to phenomena such as bias. Social locations `at the margins' that have histories of repression and resistance have better awareness of injustices and can more easily question the `naturalness' of existing power structures \cite{hooks2000feminist, d2023data, noauthor_feminist_nodate}. In the context of algorithmic bias and predictive policing, this means that those in positions of power, such as public sector actors and academics, may inadvertently reaffirm fundamentally unjust systems due to their distance from the lived experience of incarceration. 

Our findings emphasize the importance of centering the needs and experiential understanding of system-impacted communities through community-based participatory research in designing and evaluating responses to algorithmic bias. 
A critical consideration of positionality and bias may promote interventions that can serve affected communities' stated goals of community investment and healing, contributing to the development of interventions that address the root cause of algorithmic bias in predictive policing. Our results suggest that such a community-centered response to bias would in fact leave no room for the existence of crime prediction tools like predictive policing, which take the current system as a given, and would prompt a reimagination of ethical data science grounded in abolitionist philosophy and a refusal of oppressive structures \cite{benjamin_informed_2016, barabas_refusal_2022}. 

\section{Conclusion}




In this article, we presented a critical, qualitative study on evidence of enforcement bias within the Chicago crime prediction algorithm \cite{rotaru_event-level_2022}. Our research, based on interviews with various stakeholders, including community organizations, academic researchers, and public sector figures in the Chicago area, reveals that stakeholders offer diverse diagnoses of this bias, strategically leveraging these diagnoses to promote interventions that align with their respective positionality.

 These interventions encompass a range of approaches, such as using data to reform police practices that contributed to biased data collection; advocating for increased transparency and accountability in the development and utilization of predictive policing tools; discontinuing the use of such tools altogether; involving a diverse set of stakeholders in the creation and evaluation of these tools; and proposing a shift away from investing in data-driven policing reforms toward more abolitionist goals, such as community investment. We examined the implicit assumptions and the scope underlying these diverse responses to algorithmic bias, demonstrating that they are grounded in reformist and abolitionist thinking. We concluded by reflecting on the divergent AI hype views reported by participants and its relationship to lived experience with the criminal legal system. We advocate for centering the interests and experiential knowledge of system-impacted communities to ensure that evidence of algorithmic bias can be mobilized as a device to challenge the status quo. 

\section{Ethical Considerations}

\subsection{Researcher Positionality}

As white women academic researchers from Europe and the US, we acknowledge that our personal and professional backgrounds shaped how we selected research questions and analyzed data, as well as the manner in which we interacted with study participants from backgrounds similar and dissimilar to ours. Both authors have a privileged academic affiliation, race, and lack of personal contact with the US criminal legal system, which undoubtedly shaped the information that grassroots community organizations and system-impacted interviewees were comfortable sharing with us. In fact, the theme of exploitation of marginalized communities by academic researchers came up explicitly in one interview, which we discuss in \S\ref{sec:ethical_concerns}. On the other hand, our privileged position as white academic researchers from well-recognized institutions helped build rapport with other academic researchers in the study, who had similarly received training at elite academic institutions. 

\subsection{Ethical Concerns}
\label{sec:ethical_concerns}


\textbf{Privacy and anonymity.} In line with ACM Ethics Guidelines and the requirements of our IRB approval, participants' identities were protected by pseudonymizing interview transcripts. The document matching pseudonym to identifiable information was stored separately from the research data. 

\noindent \textbf{Engaging non-extractively with affected communities.} The primary ethical challenge we encountered had to do with how to involve stakeholders impacted by the criminal legal system in our research in a non-extractive and mutually beneficial manner. Due to our own resource constraints, we did not compensate interviewees for their time, which meant there was no direct benefit for study participation, and our positionality meant that affected stakeholders had little incentive or reason to place their trust in us. This became particularly salient at the end of an interview with a system-impacted participant, who works with returning citizens in Chicago. In response to a question about referring us to other study participants, they responded: ``I'm going to be completely transparent with you. Our community is exploited, often by academia. And, and so I don't feel comfortable because of that constant exploitation to put other people in that position. ... 
[Academic researchers] take our expertise, they subject us to trauma by asking us to tell our stories, and then they ... 
never come back and connect it to us.''

We acknowledge the respondent's concern and plan to share all research outputs with research participants, crediting them when desired, with the understanding that we cannot give back as much as we have received from their participation. The experience of this power differential demonstrated to us the importance of building longer-lasting community relationships outside the scope of academic projects prior to engaging in research, and increasing communities' ownership and participation of the research, with the gold standard in mind of participatory action research. 


\subsection{Adverse Impacts}

By focusing on the Chicago crime prediction algorithm, our research may inadvertently publicize the availability of the tool or promote more intensive policing. Beyond recognizing this limitation, we wish to make explicit that we do not support the development of this (or similar tools) in any form. We hope that, by shining light on the tool and the different concerns it raises, we can show how and why solutions addressed at tackling bias often do not end up benefiting the affected communities. 


	\begin{acks}
		We thank the respondents from community organizations in Chicago and all other interview participants for their time, trust and invaluable knowledge. The research was supported by the first author's affiliation with the Oxford Internet Institute’s Programme on AI, Government and Policy, funded by the Dieter Schwarz Stiftung gGmbH, and the second author's fellowship at the Berkman Klein Center for Internet \& Society at Harvard University. We thank the Embedded EthiCS group at Harvard University for their feedback on this work. 
	\end{acks}
	
\bibliographystyle{ACM-Reference-Format}
\bibliography{Law+ML}


\begin{thebibliography}{81}


\ifx \showCODEN    \undefined \def \showCODEN     #1{\unskip}     \fi
\ifx \showDOI      \undefined \def \showDOI       #1{#1}\fi
\ifx \showISBNx    \undefined \def \showISBNx     #1{\unskip}     \fi
\ifx \showISBNxiii \undefined \def \showISBNxiii  #1{\unskip}     \fi
\ifx \showISSN     \undefined \def \showISSN      #1{\unskip}     \fi
\ifx \showLCCN     \undefined \def \showLCCN      #1{\unskip}     \fi
\ifx \shownote     \undefined \def \shownote      #1{#1}          \fi
\ifx \showarticletitle \undefined \def \showarticletitle #1{#1}   \fi
\ifx \showURL      \undefined \def \showURL       {\relax}        \fi
\providecommand\bibfield[2]{#2}
\providecommand\bibinfo[2]{#2}
\providecommand\natexlab[1]{#1}
\providecommand\showeprint[2][]{arXiv:#2}

\bibitem[Abebe et~al\mbox{.}(2020)]%
        {abebe_roles_2020}
\bibfield{author}{\bibinfo{person}{Rediet Abebe}, \bibinfo{person}{Solon Barocas}, \bibinfo{person}{Jon Kleinberg}, \bibinfo{person}{Karen Levy}, \bibinfo{person}{Manish Raghavan}, {and} \bibinfo{person}{David~G. Robinson}.} \bibinfo{year}{2020}\natexlab{}.
\newblock \showarticletitle{Roles for {Computing} in {Social} {Change}}. In \bibinfo{booktitle}{\emph{Proceedings of the 2020 {Conference} on {Fairness}, {Accountability}, and {Transparency}}}. \bibinfo{pages}{252--260}.
\newblock
\urldef\tempurl%
\url{https://doi.org/10.1145/3351095.3372871}
\showDOI{\tempurl}
\newblock
\shownote{arXiv:1912.04883 [cs]}.


\bibitem[Balayn and G{\"u}rses(2021)]%
        {balayn2021beyond}
\bibfield{author}{\bibinfo{person}{Agathe Balayn} {and} \bibinfo{person}{Seda G{\"u}rses}.} \bibinfo{year}{2021}\natexlab{}.
\newblock \showarticletitle{Beyond Debiasing: Regulating AI and its inequalities}.
\newblock \bibinfo{journal}{\emph{EDRi Report. https://edri. org/wp-content/uploads/2021/09/EDRi\_Beyond-Debiasing-Report\_Online. pdf}} (\bibinfo{year}{2021}).
\newblock


\bibitem[Barabas(2019)]%
        {barabas_beyond_2019}
\bibfield{author}{\bibinfo{person}{Chelsea Barabas}.} \bibinfo{year}{2019}\natexlab{}.
\newblock \bibinfo{booktitle}{\emph{Beyond {Bias}: {Re}-{Imagining} the {Terms} of ‘{Ethical} {AI}’ in {Criminal} {Law}}}.
\newblock \bibinfo{type}{{SSRN} {Scholarly} {Paper}} ID 3377921. \bibinfo{institution}{Social Science Research Network}, \bibinfo{address}{Rochester, NY}.
\newblock
\urldef\tempurl%
\url{https://doi.org/10.2139/ssrn.3377921}
\showDOI{\tempurl}


\bibitem[Barabas(2022)]%
        {barabas_refusal_2022}
\bibfield{author}{\bibinfo{person}{Chelsea Barabas}.} \bibinfo{year}{2022}\natexlab{}.
\newblock \showarticletitle{Refusal in {Data} {Ethics}: {Re}-{Imagining} the {Code} {Beneath} the {Code} of {Computation} in the {Carceral} {State}}.
\newblock \bibinfo{journal}{\emph{Engaging Science, Technology, and Society}} \bibinfo{volume}{8}, \bibinfo{number}{2} (\bibinfo{date}{Sept.} \bibinfo{year}{2022}), \bibinfo{pages}{35--57}.
\newblock
\showISSN{2413-8053}
\urldef\tempurl%
\url{https://doi.org/10.17351/ests2022.1233}
\showDOI{\tempurl}
\newblock
\shownote{Number: 2}.


\bibitem[Barabas et~al\mbox{.}(2018)]%
        {barabas_interventions_2018}
\bibfield{author}{\bibinfo{person}{Chelsea Barabas}, \bibinfo{person}{Madars Virza}, \bibinfo{person}{Karthik Dinakar}, \bibinfo{person}{Joichi Ito}, {and} \bibinfo{person}{Jonathan Zittrain}.} \bibinfo{year}{2018}\natexlab{}.
\newblock \showarticletitle{Interventions over {Predictions}: {Reframing} the {Ethical} {Debate} for {Actuarial} {Risk} {Assessment}}. In \bibinfo{booktitle}{\emph{Conference on {Fairness}, {Accountability} and {Transparency}}}. \bibinfo{pages}{62--76}.
\newblock
\urldef\tempurl%
\url{http://proceedings.mlr.press/v81/barabas18a.html}
\showURL{%
\tempurl}


\bibitem[Barba and Baustin(2023)]%
        {barba_are_2023}
\bibfield{author}{\bibinfo{person}{Michael Barba} {and} \bibinfo{person}{Noah Baustin}.} \bibinfo{year}{2023}\natexlab{}.
\newblock \showarticletitle{Are {San} {Francisco} {Police} {Officers} {Misreporting} the {Races} of {People} {They} {Stop}?}
\newblock \bibinfo{journal}{\emph{The San Francisco Standard}} (\bibinfo{date}{Sept.} \bibinfo{year}{2023}).
\newblock
\urldef\tempurl%
\url{https://sfstandard.com/2023/09/13/san-francisco-police-officer-misrepresented-race-bias-investigation/}
\showURL{%
\tempurl}


\bibitem[Benjamin(2016)]%
        {benjamin_informed_2016}
\bibfield{author}{\bibinfo{person}{Ruha Benjamin}.} \bibinfo{year}{2016}\natexlab{}.
\newblock \showarticletitle{Informed {Refusal}: {Toward} a {Justice}-based {Bioethics}}.
\newblock \bibinfo{journal}{\emph{Science, Technology, \& Human Values}} \bibinfo{volume}{41}, \bibinfo{number}{6} (\bibinfo{date}{Nov.} \bibinfo{year}{2016}), \bibinfo{pages}{967--990}.
\newblock
\showISSN{0162-2439}
\urldef\tempurl%
\url{https://doi.org/10.1177/0162243916656059}
\showDOI{\tempurl}
\newblock
\shownote{Publisher: SAGE Publications Inc}.


\bibitem[Benjamin(2019)]%
        {benjamin_race_2019}
\bibfield{author}{\bibinfo{person}{Ruha Benjamin}.} \bibinfo{year}{2019}\natexlab{}.
\newblock \bibinfo{booktitle}{\emph{Race {After} {Technology}: {Abolitionist} {Tools} for the {New} {Jim} {Code}}}.
\newblock \bibinfo{publisher}{John Wiley \& Sons}.
\newblock
\showISBNx{978-1-5095-2643-7}


\bibitem[Bowell({[n.\,d.]})]%
        {noauthor_feminist_nodate}
\bibfield{author}{\bibinfo{person}{T Bowell}.} \bibinfo{year}{[n.\,d.]}\natexlab{}.
\newblock \bibinfo{title}{Feminist {Standpoint} {Theory} {\textbar} {Internet} {Encyclopedia} of {Philosophy}}.
\newblock
\newblock
\urldef\tempurl%
\url{https://iep.utm.edu/fem-stan/}
\showURL{%
\tempurl}


\bibitem[Brayne(2020)]%
        {brayne_predict_2020}
\bibfield{author}{\bibinfo{person}{Sarah Brayne}.} \bibinfo{year}{2020}\natexlab{}.
\newblock \bibinfo{booktitle}{\emph{Predict and {Surveil}: {Data}, {Discretion}, and the {Future} of {Policing}}}.
\newblock \bibinfo{publisher}{Oxford University Press}.
\newblock
\showISBNx{978-0-19-068409-9}


\bibitem[Broussard et~al\mbox{.}(2019)]%
        {broussard_artificial_2019}
\bibfield{author}{\bibinfo{person}{Meredith Broussard}, \bibinfo{person}{Nicholas Diakopoulos}, \bibinfo{person}{Andrea~L. Guzman}, \bibinfo{person}{Rediet Abebe}, \bibinfo{person}{Michel Dupagne}, {and} \bibinfo{person}{Ching-Hua Chuan}.} \bibinfo{year}{2019}\natexlab{}.
\newblock \showarticletitle{Artificial {Intelligence} and {Journalism}}.
\newblock \bibinfo{journal}{\emph{Journalism \& Mass Communication Quarterly}} \bibinfo{volume}{96}, \bibinfo{number}{3} (\bibinfo{date}{Sept.} \bibinfo{year}{2019}), \bibinfo{pages}{673--695}.
\newblock
\showISSN{1077-6990}
\urldef\tempurl%
\url{https://doi.org/10.1177/1077699019859901}
\showDOI{\tempurl}
\newblock
\shownote{Publisher: SAGE Publications Inc}.


\bibitem[Bruno et~al\mbox{.}(2014)]%
        {bruno_statactivism_2014}
\bibfield{author}{\bibinfo{person}{Isabelle Bruno}, \bibinfo{person}{Emmanuel Didier}, {and} \bibinfo{person}{Tommaso Vitale}.} \bibinfo{year}{2014}\natexlab{}.
\newblock \bibinfo{title}{Statactivism: {Forms} of {Action} between {Disclosure} and {Affirmation}}.
\newblock
\newblock
\urldef\tempurl%
\url{https://papers.ssrn.com/abstract=2466882}
\showURL{%
\tempurl}


\bibitem[Desrosi{\`e}res(1998)]%
        {desrosieres1998politics}
\bibfield{author}{\bibinfo{person}{Alain Desrosi{\`e}res}.} \bibinfo{year}{1998}\natexlab{}.
\newblock \bibinfo{booktitle}{\emph{The politics of large numbers: A history of statistical reasoning}}.
\newblock \bibinfo{publisher}{Harvard University Press}.
\newblock


\bibitem[D'Ignazio(2024)]%
        {dignazio_counting_2024}
\bibfield{author}{\bibinfo{person}{Catherine D'Ignazio}.} \bibinfo{year}{2024}\natexlab{}.
\newblock \bibinfo{booktitle}{\emph{Counting {Feminicide}: {Data} {Feminism} in {Action}}}.
\newblock \bibinfo{publisher}{The MIT Press}, \bibinfo{address}{Cambridge, Massachusetts}.
\newblock


\bibitem[D'ignazio and Klein(2023)]%
        {d2023data}
\bibfield{author}{\bibinfo{person}{Catherine D'ignazio} {and} \bibinfo{person}{Lauren~F Klein}.} \bibinfo{year}{2023}\natexlab{}.
\newblock \bibinfo{booktitle}{\emph{Data feminism}}.
\newblock \bibinfo{publisher}{MIT press}.
\newblock


\bibitem[Edwards and Holland(2013)]%
        {edwards_what_2013}
\bibfield{author}{\bibinfo{person}{Rosalind Edwards} {and} \bibinfo{person}{Janet Holland}.} \bibinfo{year}{2013}\natexlab{}.
\newblock \bibinfo{booktitle}{\emph{What is {Qualitative} {Interviewing}?}}
\newblock \bibinfo{publisher}{Bloomsbury Academic}.
\newblock
\showISBNx{978-1-84966-802-6 978-1-84966-801-9}
\urldef\tempurl%
\url{https://doi.org/10.5040/9781472545244}
\showDOI{\tempurl}
\newblock
\shownote{Accepted: 2022-10-14T14:53:13Z}.


\bibitem[Ensign et~al\mbox{.}(2018)]%
        {ensign_runaway_2018}
\bibfield{author}{\bibinfo{person}{Danielle Ensign}, \bibinfo{person}{Sorelle Friedler}, \bibinfo{person}{Scott Neville}, \bibinfo{person}{Carlos Scheidegger}, {and} \bibinfo{person}{Suresh Venkatasubramanian}.} \bibinfo{year}{2018}\natexlab{}.
\newblock \showarticletitle{Runaway {Feedback} {Loops} in {Predictive} {Policing}}.
\newblock \bibinfo{journal}{\emph{Proceedings of Machine Learning Research}}  \bibinfo{volume}{81} (\bibinfo{date}{Jan.} \bibinfo{year}{2018}).
\newblock
\urldef\tempurl%
\url{https://scholarship.haverford.edu/compsci_facpubs/113}
\showURL{%
\tempurl}


\bibitem[Espeland and Vannebo(2007)]%
        {espeland_accountability_2007}
\bibfield{author}{\bibinfo{person}{Wendy~Nelson Espeland} {and} \bibinfo{person}{Berit~Irene Vannebo}.} \bibinfo{year}{2007}\natexlab{}.
\newblock \showarticletitle{Accountability, quantification, and law}.
\newblock In \bibinfo{booktitle}{\emph{Annual {Review} of {Law} and {Social} {Science}}}, \bibfield{editor}{\bibinfo{person}{John Hagan}, \bibinfo{person}{Kim~Lane Scheppele}, {and} \bibinfo{person}{Tom Tyler}} (Eds.). \bibinfo{pages}{21--43}.
\newblock
\showISBNx{978-0-8243-4103-9}
\urldef\tempurl%
\url{https://doi.org/10.1146/annurev.lawsocsci.2.081805.105908}
\showDOI{\tempurl}


\bibitem[Eubanks(2018)]%
        {eubanks_automating_2018}
\bibfield{author}{\bibinfo{person}{Virginia Eubanks}.} \bibinfo{year}{2018}\natexlab{}.
\newblock \bibinfo{booktitle}{\emph{Automating {Inequality}: {How} {High}-{Tech} {Tools} {Profile}, {Police}, and {Punish} the {Poor}}}.
\newblock \bibinfo{publisher}{St. Martin's Publishing Group}.
\newblock
\showISBNx{978-1-4668-8596-7}


\bibitem[Fahse et~al\mbox{.}(2021)]%
        {fahse_managing_2021}
\bibfield{author}{\bibinfo{person}{Tobias Fahse}, \bibinfo{person}{Viktoria Huber}, {and} \bibinfo{person}{Benjamin van Giffen}.} \bibinfo{year}{2021}\natexlab{}.
\newblock \showarticletitle{Managing {Bias} in {Machine} {Learning} {Projects}}. In \bibinfo{booktitle}{\emph{Innovation {Through} {Information} {Systems}}} \emph{(\bibinfo{series}{Lecture {Notes} in {Information} {Systems} and {Organisation}})}, \bibfield{editor}{\bibinfo{person}{Frederik Ahlemann}, \bibinfo{person}{Reinhard Schütte}, {and} \bibinfo{person}{Stefan Stieglitz}} (Eds.). \bibinfo{publisher}{Springer International Publishing}, \bibinfo{address}{Cham}, \bibinfo{pages}{94--109}.
\newblock
\showISBNx{978-3-030-86797-3}
\urldef\tempurl%
\url{https://doi.org/10.1007/978-3-030-86797-3_7}
\showDOI{\tempurl}


\bibitem[Fazelpour and Danks(2021)]%
        {fazelpour_algorithmic_2021}
\bibfield{author}{\bibinfo{person}{Sina Fazelpour} {and} \bibinfo{person}{David Danks}.} \bibinfo{year}{2021}\natexlab{}.
\newblock \showarticletitle{Algorithmic bias: {Senses}, sources, solutions}.
\newblock \bibinfo{journal}{\emph{Philosophy Compass}} \bibinfo{volume}{16}, \bibinfo{number}{8} (\bibinfo{year}{2021}), \bibinfo{pages}{e12760}.
\newblock
\showISSN{1747-9991}
\urldef\tempurl%
\url{https://doi.org/10.1111/phc3.12760}
\showDOI{\tempurl}
\newblock
\shownote{\_eprint: https://onlinelibrary.wiley.com/doi/pdf/10.1111/phc3.12760}.


\bibitem[Feathers(2021)]%
        {feathers_gunshot-detecting_2021}
\bibfield{author}{\bibinfo{person}{Todd Feathers}.} \bibinfo{year}{2021}\natexlab{}.
\newblock \bibinfo{title}{Gunshot-{Detecting} {Tech} {Is} {Summoning} {Armed} {Police} to {Black} {Neighborhoods}}.
\newblock
\newblock
\urldef\tempurl%
\url{https://www.vice.com/en/article/88nd3z/gunshot-detecting-tech-is-summoning-armed-police-to-black-neighborhoods}
\showURL{%
\tempurl}


\bibitem[Ferguson and Witzburg(2021)]%
        {ferguson2021chicago}
\bibfield{author}{\bibinfo{person}{J Ferguson} {and} \bibinfo{person}{Deborah Witzburg}.} \bibinfo{year}{2021}\natexlab{}.
\newblock \showarticletitle{The Chicago Police Department’s Use of Shotspotter Technology}.
\newblock \bibinfo{journal}{\emph{Chicago, IL: City of Chicago Office of Inspector General}} (\bibinfo{year}{2021}).
\newblock


\bibitem[Franchi et~al\mbox{.}(2023)]%
        {franchi_detecting_2023}
\bibfield{author}{\bibinfo{person}{Matt Franchi}, \bibinfo{person}{J.D. Zamfirescu-Pereira}, \bibinfo{person}{Wendy Ju}, {and} \bibinfo{person}{Emma Pierson}.} \bibinfo{year}{2023}\natexlab{}.
\newblock \showarticletitle{Detecting disparities in police deployments using dashcam data}. In \bibinfo{booktitle}{\emph{Proceedings of the 2023 {ACM} {Conference} on {Fairness}, {Accountability}, and {Transparency}}} \emph{(\bibinfo{series}{{FAccT} '23})}. \bibinfo{publisher}{Association for Computing Machinery}, \bibinfo{address}{New York, NY, USA}, \bibinfo{pages}{534--544}.
\newblock
\showISBNx{9798400701924}
\urldef\tempurl%
\url{https://doi.org/10.1145/3593013.3594020}
\showDOI{\tempurl}


\bibitem[Frey(2018)]%
        {frey_sage_2018}
\bibfield{author}{\bibinfo{person}{Bruce~B. Frey}.} \bibinfo{year}{2018}\natexlab{}.
\newblock \bibinfo{booktitle}{\emph{The {SAGE} {Encyclopedia} of {Educational} {Research}, {Measurement}, and {Evaluation}}}.
\newblock \bibinfo{publisher}{SAGE Publications, Inc.}
\newblock
\showISBNx{978-1-5063-2613-9}
\urldef\tempurl%
\url{https://doi.org/10.4135/9781506326139}
\showDOI{\tempurl}


\bibitem[Gebru(2019)]%
        {gebru_oxford_2019}
\bibfield{author}{\bibinfo{person}{Timnit Gebru}.} \bibinfo{year}{2019}\natexlab{}.
\newblock \bibinfo{title}{Oxford {Handbook} on {AI} {Ethics} {Book} {Chapter} on {Race} and {Gender}}.
\newblock
\newblock
\urldef\tempurl%
\url{https://doi.org/10.48550/arXiv.1908.06165}
\showDOI{\tempurl}
\newblock
\shownote{arXiv:1908.06165 [cs]}.


\bibitem[Gorz(1968)]%
        {gorz1968strategy}
\bibfield{author}{\bibinfo{person}{Andre Gorz}.} \bibinfo{year}{1968}\natexlab{}.
\newblock \showarticletitle{Strategy for labor. A radical proposal}.
\newblock \bibinfo{journal}{\emph{Science and Society}} \bibinfo{volume}{32}, \bibinfo{number}{4} (\bibinfo{year}{1968}).
\newblock


\bibitem[Green(2019)]%
        {green2019good}
\bibfield{author}{\bibinfo{person}{Ben Green}.} \bibinfo{year}{2019}\natexlab{}.
\newblock \showarticletitle{Good” isn’t good enough}. In \bibinfo{booktitle}{\emph{Proceedings of the AI for Social Good workshop at NeurIPS}}, Vol.~\bibinfo{volume}{17}.
\newblock


\bibitem[Green(2021)]%
        {green_data_2021}
\bibfield{author}{\bibinfo{person}{Ben Green}.} \bibinfo{year}{2021}\natexlab{}.
\newblock \showarticletitle{Data {Science} as {Political} {Action}: {Grounding} {Data} {Science} in a {Politics} of {Justice}}.
\newblock \bibinfo{journal}{\emph{Journal of Social Computing}} \bibinfo{volume}{2}, \bibinfo{number}{3} (\bibinfo{date}{Sept.} \bibinfo{year}{2021}), \bibinfo{pages}{249--265}.
\newblock
\showISSN{2688-5255}
\urldef\tempurl%
\url{https://doi.org/10.23919/JSC.2021.0029}
\showDOI{\tempurl}
\newblock
\shownote{arXiv:1811.03435 [cs]}.


\bibitem[Green(2022)]%
        {green_escaping_2022}
\bibfield{author}{\bibinfo{person}{Ben Green}.} \bibinfo{year}{2022}\natexlab{}.
\newblock \showarticletitle{Escaping the {Impossibility} of {Fairness}: {From} {Formal} to {Substantive} {Algorithmic} {Fairness}}.
\newblock \bibinfo{journal}{\emph{Philosophy \& Technology}} \bibinfo{volume}{35}, \bibinfo{number}{4} (\bibinfo{date}{Oct.} \bibinfo{year}{2022}), \bibinfo{pages}{90}.
\newblock
\showISSN{2210-5441}
\urldef\tempurl%
\url{https://doi.org/10.1007/s13347-022-00584-6}
\showDOI{\tempurl}


\bibitem[Green and Viljoen(2020)]%
        {green_algorithmic_2020}
\bibfield{author}{\bibinfo{person}{Ben Green} {and} \bibinfo{person}{Salomé Viljoen}.} \bibinfo{year}{2020}\natexlab{}.
\newblock \showarticletitle{Algorithmic realism: expanding the boundaries of algorithmic thought}. In \bibinfo{booktitle}{\emph{Proceedings of the 2020 {Conference} on {Fairness}, {Accountability}, and {Transparency}}} \emph{(\bibinfo{series}{{FAT}* '20})}. \bibinfo{publisher}{Association for Computing Machinery}, \bibinfo{address}{Barcelona, Spain}, \bibinfo{pages}{19--31}.
\newblock
\showISBNx{978-1-4503-6936-7}
\urldef\tempurl%
\url{https://doi.org/10.1145/3351095.3372840}
\showDOI{\tempurl}


\bibitem[Grill et~al\mbox{.}(2023)]%
        {grill_bias_2023}
\bibfield{author}{\bibinfo{person}{Gabriel Grill}, \bibinfo{person}{Fabian Fischer}, {and} \bibinfo{person}{Florian Cech}.} \bibinfo{year}{2023}\natexlab{}.
\newblock \showarticletitle{Bias as {Boundary} {Object}: {Unpacking} {The} {Politics} {Of} {An} {Austerity} {Algorithm} {Using} {Bias} {Frameworks}}. In \bibinfo{booktitle}{\emph{Proceedings of the 2023 {ACM} {Conference} on {Fairness}, {Accountability}, and {Transparency}}} \emph{(\bibinfo{series}{{FAccT} '23})}. \bibinfo{publisher}{Association for Computing Machinery}, \bibinfo{address}{New York, NY, USA}, \bibinfo{pages}{1838--1849}.
\newblock
\showISBNx{9798400701924}
\urldef\tempurl%
\url{https://doi.org/10.1145/3593013.3594120}
\showDOI{\tempurl}


\bibitem[Harcourt(2008)]%
        {harcourt_against_2008}
\bibfield{author}{\bibinfo{person}{Bernard~E. Harcourt}.} \bibinfo{year}{2008}\natexlab{}.
\newblock \bibinfo{booktitle}{\emph{Against {Prediction}: {Profiling}, {Policing}, and {Punishing} in an {Actuarial} {Age}}}.
\newblock \bibinfo{publisher}{University of Chicago Press}.
\newblock
\showISBNx{978-0-226-31599-7}


\bibitem[Harding(1992)]%
        {harding_rethinking_1992}
\bibfield{author}{\bibinfo{person}{Sandra Harding}.} \bibinfo{year}{1992}\natexlab{}.
\newblock \showarticletitle{Rethinking {Standpoint} {Epistemology}: {What} is ``{Strong} {Objectivity}?"}.
\newblock \bibinfo{journal}{\emph{The Centennial Review}} \bibinfo{volume}{36}, \bibinfo{number}{3} (\bibinfo{year}{1992}), \bibinfo{pages}{437--470}.
\newblock
\showISSN{0162-0177}
\urldef\tempurl%
\url{https://www.jstor.org/stable/23739232}
\showURL{%
\tempurl}


\bibitem[Hill(2023)]%
        {hill2023your}
\bibfield{author}{\bibinfo{person}{Kashmir Hill}.} \bibinfo{year}{2023}\natexlab{}.
\newblock \bibinfo{booktitle}{\emph{Your Face Belongs to Us: The Secretive Startup Dismantling Your Privacy}}.
\newblock \bibinfo{publisher}{Simon and Schuster}.
\newblock


\bibitem[Hoffmann(2019)]%
        {hoffmann2019fairness}
\bibfield{author}{\bibinfo{person}{Anna~Lauren Hoffmann}.} \bibinfo{year}{2019}\natexlab{}.
\newblock \showarticletitle{Where fairness fails: data, algorithms, and the limits of antidiscrimination discourse}.
\newblock \bibinfo{journal}{\emph{Information, Communication \& Society}} \bibinfo{volume}{22}, \bibinfo{number}{7} (\bibinfo{year}{2019}), \bibinfo{pages}{900--915}.
\newblock


\bibitem[Hooks(2000)]%
        {hooks2000feminist}
\bibfield{author}{\bibinfo{person}{Bell Hooks}.} \bibinfo{year}{2000}\natexlab{}.
\newblock \bibinfo{booktitle}{\emph{Feminist theory: From margin to center}}.
\newblock \bibinfo{publisher}{Pluto Press}.
\newblock


\bibitem[Karakatsanis(2019)]%
        {karakatsanis2019usual}
\bibfield{author}{\bibinfo{person}{Alec Karakatsanis}.} \bibinfo{year}{2019}\natexlab{}.
\newblock \bibinfo{booktitle}{\emph{Usual cruelty: The complicity of lawyers in the criminal injustice system}}.
\newblock \bibinfo{publisher}{The New Press}.
\newblock


\bibitem[Karppinen and Moe(2011)]%
        {karppinen_what_2011}
\bibfield{author}{\bibinfo{person}{Kari Karppinen} {and} \bibinfo{person}{Hallvard Moe}.} \bibinfo{year}{2011}\natexlab{}.
\newblock \showarticletitle{What we talk about when we talk about document analysis}.
\newblock In \bibinfo{booktitle}{\emph{Trends in {Communication} {Policy} {Research}}}. \bibinfo{publisher}{Intellect}, \bibinfo{address}{Bristol}.
\newblock
\showISBNx{978-1-84150-467-4}


\bibitem[Kennedy et~al\mbox{.}(2018)]%
        {kennedy_risk-based_2018}
\bibfield{author}{\bibinfo{person}{Leslie~W. Kennedy}, \bibinfo{person}{Joel~M. Caplan}, {and} \bibinfo{person}{Eric~L. Piza}.} \bibinfo{year}{2018}\natexlab{}.
\newblock \bibinfo{booktitle}{\emph{Risk-{Based} {Policing}: {Evidence}-{Based} {Crime} {Prevention} with {Big} {Data} and {Spatial} {Analytics}}}.
\newblock \bibinfo{publisher}{Univ of California Press}.
\newblock
\showISBNx{978-0-520-29563-6}


\bibitem[Kunichoff and Sier(2017)]%
        {kunichoff_contradictions_2017}
\bibfield{author}{\bibinfo{person}{Yana Kunichoff} {and} \bibinfo{person}{Patrick Sier}.} \bibinfo{year}{2017}\natexlab{}.
\newblock \bibinfo{title}{The {Contradictions} of {Chicago} {Police}'s {Secretive} {List}}.
\newblock
\newblock
\urldef\tempurl%
\url{https://www.chicagomag.com/city-life/August-2017/Chicago-Police-Strategic-Subject-List/}
\showURL{%
\tempurl}


\bibitem[Lipton(1977)]%
        {lipton_why_1977}
\bibfield{author}{\bibinfo{person}{Michael Lipton}.} \bibinfo{year}{1977}\natexlab{}.
\newblock \bibinfo{booktitle}{\emph{Why poor people stay poor : a study of urban bias in world development}}.
\newblock \bibinfo{publisher}{Temple Smith ; Australian National University Press}.
\newblock
\urldef\tempurl%
\url{https://openresearch-repository.anu.edu.au/handle/1885/114902}
\showURL{%
\tempurl}
\newblock
\shownote{Accepted: 2017-04-18T05:59:25Z Last Modified: 2022-11-17}.


\bibitem[Lum and Isaac(2016)]%
        {lum_predict_2016}
\bibfield{author}{\bibinfo{person}{Kristian Lum} {and} \bibinfo{person}{William Isaac}.} \bibinfo{year}{2016}\natexlab{}.
\newblock \showarticletitle{To predict and serve?}
\newblock \bibinfo{journal}{\emph{Significance}} \bibinfo{volume}{13}, \bibinfo{number}{5} (\bibinfo{year}{2016}), \bibinfo{pages}{14--19}.
\newblock
\showISSN{1740-9713}
\urldef\tempurl%
\url{https://doi.org/10.1111/j.1740-9713.2016.00960.x}
\showDOI{\tempurl}
\newblock
\shownote{\_eprint: https://onlinelibrary.wiley.com/doi/pdf/10.1111/j.1740-9713.2016.00960.x}.


\bibitem[Mann et~al\mbox{.}(2003)]%
        {mann2003sousveillance}
\bibfield{author}{\bibinfo{person}{Steve Mann}, \bibinfo{person}{Jason Nolan}, {and} \bibinfo{person}{Barry Wellman}.} \bibinfo{year}{2003}\natexlab{}.
\newblock \showarticletitle{Sousveillance: Inventing and using wearable computing devices for data collection in surveillance environments.}
\newblock \bibinfo{journal}{\emph{Surveillance \& society}} \bibinfo{volume}{1}, \bibinfo{number}{3} (\bibinfo{year}{2003}), \bibinfo{pages}{331--355}.
\newblock


\bibitem[Mayson({[n.\,d.]})]%
        {mayson_bias_nodate}
\bibfield{author}{\bibinfo{person}{Sandra~G. Mayson}.} \bibinfo{year}{[n.\,d.]}\natexlab{}.
\newblock \bibinfo{title}{Bias {In}, {Bias} {Out}}.
\newblock
\newblock
\urldef\tempurl%
\url{https://www.yalelawjournal.org/article/bias-in-bias-out}
\showURL{%
\tempurl}


\bibitem[McLeod(2018)]%
        {mcleod2018envisioning}
\bibfield{author}{\bibinfo{person}{Allegra~M McLeod}.} \bibinfo{year}{2018}\natexlab{}.
\newblock \showarticletitle{Envisioning abolition democracy}.
\newblock \bibinfo{journal}{\emph{Harv. L. Rev.}}  \bibinfo{volume}{132} (\bibinfo{year}{2018}), \bibinfo{pages}{1613}.
\newblock


\bibitem[Meyer and Graybill(2016)]%
        {meyer_suburban_2016}
\bibfield{author}{\bibinfo{person}{William~B. Meyer} {and} \bibinfo{person}{Jessica~K. Graybill}.} \bibinfo{year}{2016}\natexlab{}.
\newblock \showarticletitle{The suburban bias of {American} society?}
\newblock \bibinfo{journal}{\emph{Urban Geography}} \bibinfo{volume}{37}, \bibinfo{number}{6} (\bibinfo{date}{Aug.} \bibinfo{year}{2016}), \bibinfo{pages}{863--882}.
\newblock
\showISSN{0272-3638}
\urldef\tempurl%
\url{https://doi.org/10.1080/02723638.2015.1118990}
\showDOI{\tempurl}
\newblock
\shownote{Publisher: Routledge \_eprint: https://doi.org/10.1080/02723638.2015.1118990}.


\bibitem[Miles et~al\mbox{.}(2014)]%
        {miles_qualitative_2014}
\bibfield{author}{\bibinfo{person}{Matthew~B. Miles}, \bibinfo{person}{A.~Michael Huberman}, {and} \bibinfo{person}{Johnny Saldana}.} \bibinfo{year}{2014}\natexlab{}.
\newblock \bibinfo{booktitle}{\emph{Qualitative {Data} {Analysis}}}.
\newblock \bibinfo{publisher}{SAGE}.
\newblock
\showISBNx{978-1-4522-5787-7}


\bibitem[Mitchell et~al\mbox{.}(2019)]%
        {mitchell_model_2019}
\bibfield{author}{\bibinfo{person}{Margaret Mitchell}, \bibinfo{person}{Simone Wu}, \bibinfo{person}{Andrew Zaldivar}, \bibinfo{person}{Parker Barnes}, \bibinfo{person}{Lucy Vasserman}, \bibinfo{person}{Ben Hutchinson}, \bibinfo{person}{Elena Spitzer}, \bibinfo{person}{Inioluwa~Deborah Raji}, {and} \bibinfo{person}{Timnit Gebru}.} \bibinfo{year}{2019}\natexlab{}.
\newblock \showarticletitle{Model {Cards} for {Model} {Reporting}}. In \bibinfo{booktitle}{\emph{Proceedings of the {Conference} on {Fairness}, {Accountability}, and {Transparency}}} \emph{(\bibinfo{series}{{FAT}* '19})}. \bibinfo{publisher}{Association for Computing Machinery}, \bibinfo{address}{New York, NY, USA}, \bibinfo{pages}{220--229}.
\newblock
\showISBNx{978-1-4503-6125-5}
\urldef\tempurl%
\url{https://doi.org/10.1145/3287560.3287596}
\showDOI{\tempurl}


\bibitem[Mohtasham(2024)]%
        {mohtasham_chicago_2024}
\bibfield{author}{\bibinfo{person}{Diba Mohtasham}.} \bibinfo{year}{2024}\natexlab{}.
\newblock \showarticletitle{Chicago will drop controversial {ShotSpotter} gunfire detection system}.
\newblock \bibinfo{journal}{\emph{NPR}} (\bibinfo{date}{Feb.} \bibinfo{year}{2024}).
\newblock
\urldef\tempurl%
\url{https://www.npr.org/2024/02/15/1231394334/shotspotter-gunfire-detection-chicago-mayor-dropping}
\showURL{%
\tempurl}


\bibitem[Newell(2020)]%
        {newell_introduction_2020}
\bibfield{author}{\bibinfo{person}{Bryce Newell}.} \bibinfo{year}{2020}\natexlab{}.
\newblock \showarticletitle{Introduction: {The} {State} of {Sousveillance}}.
\newblock \bibinfo{journal}{\emph{Surveillance \& Society}} \bibinfo{volume}{18}, \bibinfo{number}{2} (\bibinfo{date}{June} \bibinfo{year}{2020}), \bibinfo{pages}{257--261}.
\newblock
\showISSN{1477-7487}
\urldef\tempurl%
\url{https://doi.org/10.24908/ss.v18i2.14013}
\showDOI{\tempurl}


\bibitem[Nierenberg(2023)]%
        {nierenberg_over_2023}
\bibfield{author}{\bibinfo{person}{Amelia Nierenberg}.} \bibinfo{year}{2023}\natexlab{}.
\newblock \showarticletitle{Over 100 {Connecticut} {State} {Troopers} {Accused} of {Faking} {Traffic} {Stops}}.
\newblock \bibinfo{journal}{\emph{The New York Times}} (\bibinfo{date}{Sept.} \bibinfo{year}{2023}).
\newblock
\showISSN{0362-4331}
\urldef\tempurl%
\url{https://www.nytimes.com/2023/09/04/nyregion/connecticut-false-tickets.html}
\showURL{%
\tempurl}


\bibitem[Parker et~al\mbox{.}(2019)]%
        {parker_snowball_2019}
\bibfield{author}{\bibinfo{person}{C. Parker}, \bibinfo{person}{S. Scott}, {and} \bibinfo{person}{A. Geddes}.} \bibinfo{year}{2019}\natexlab{}.
\newblock \showarticletitle{Snowball {Sampling}}.
\newblock \bibinfo{journal}{\emph{SAGE Research Methods Foundations}} (\bibinfo{date}{Sept.} \bibinfo{year}{2019}).
\newblock
\urldef\tempurl%
\url{http://methods.sagepub.com/foundations/snowball-sampling}
\showURL{%
\tempurl}
\newblock
\shownote{Publisher: SAGE}.


\bibitem[Passi and Barocas(2019)]%
        {passi2019problem}
\bibfield{author}{\bibinfo{person}{Samir Passi} {and} \bibinfo{person}{Solon Barocas}.} \bibinfo{year}{2019}\natexlab{}.
\newblock \showarticletitle{Problem formulation and fairness}. In \bibinfo{booktitle}{\emph{Proceedings of the conference on fairness, accountability, and transparency}}. \bibinfo{pages}{39--48}.
\newblock


\bibitem[Pearsall(2010)]%
        {pearsall2010predictive}
\bibfield{author}{\bibinfo{person}{Beth Pearsall}.} \bibinfo{year}{2010}\natexlab{}.
\newblock \showarticletitle{Predictive policing: The future of law enforcement}.
\newblock \bibinfo{journal}{\emph{National Institute of Justice Journal}} \bibinfo{volume}{266}, \bibinfo{number}{1} (\bibinfo{year}{2010}), \bibinfo{pages}{16--19}.
\newblock


\bibitem[Porter(1995)]%
        {porter_trust_1995}
\bibfield{author}{\bibinfo{person}{Theodore~M. Porter}.} \bibinfo{year}{1995}\natexlab{}.
\newblock \bibinfo{booktitle}{\emph{Trust in numbers: the pursuit of objectivity in science and public life}}.
\newblock \bibinfo{publisher}{Princeton University Press}, \bibinfo{address}{Princeton, N.J}.
\newblock
\showISBNx{978-0-691-03776-9}


\bibitem[Pruss(2023)]%
        {pruss2023ghosting}
\bibfield{author}{\bibinfo{person}{Dasha Pruss}.} \bibinfo{year}{2023}\natexlab{}.
\newblock \showarticletitle{Ghosting the Machine: Judicial Resistance to a Recidivism Risk Assessment Instrument}. In \bibinfo{booktitle}{\emph{Proceedings of the 2023 ACM Conference on Fairness, Accountability, and Transparency}}. \bibinfo{pages}{312--323}.
\newblock


\bibitem[Ricaurte(2019)]%
        {ricaurte2019data}
\bibfield{author}{\bibinfo{person}{Paola Ricaurte}.} \bibinfo{year}{2019}\natexlab{}.
\newblock \showarticletitle{Data epistemologies, the coloniality of power, and resistance}.
\newblock \bibinfo{journal}{\emph{Television \& New Media}} \bibinfo{volume}{20}, \bibinfo{number}{4} (\bibinfo{year}{2019}), \bibinfo{pages}{350--365}.
\newblock


\bibitem[Richards(1996)]%
        {richards_elite_1996}
\bibfield{author}{\bibinfo{person}{David Richards}.} \bibinfo{year}{1996}\natexlab{}.
\newblock \showarticletitle{Elite {Interviewing}: {Approaches} and {Pitfalls}}.
\newblock \bibinfo{journal}{\emph{Politics}} \bibinfo{volume}{16}, \bibinfo{number}{3} (\bibinfo{year}{1996}), \bibinfo{pages}{199--204}.
\newblock
\showISSN{1467-9256}
\urldef\tempurl%
\url{https://doi.org/10.1111/j.1467-9256.1996.tb00039.x}
\showDOI{\tempurl}
\newblock
\shownote{\_eprint: https://onlinelibrary.wiley.com/doi/pdf/10.1111/j.1467-9256.1996.tb00039.x}.


\bibitem[Richardson et~al\mbox{.}(2019)]%
        {richardson_dirty_2019}
\bibfield{author}{\bibinfo{person}{Rashida Richardson}, \bibinfo{person}{Jason Schultz}, {and} \bibinfo{person}{Kate Crawford}.} \bibinfo{year}{2019}\natexlab{}.
\newblock \bibinfo{booktitle}{\emph{Dirty {Data}, {Bad} {Predictions}: {How} {Civil} {Rights} {Violations} {Impact} {Police} {Data}, {Predictive} {Policing} {Systems}, and {Justice}}}.
\newblock \bibinfo{type}{{SSRN} {Scholarly} {Paper}} ID 3333423. \bibinfo{institution}{Social Science Research Network}, \bibinfo{address}{Rochester, NY}.
\newblock
\urldef\tempurl%
\url{https://papers.ssrn.com/abstract=3333423}
\showURL{%
\tempurl}


\bibitem[Rotaru et~al\mbox{.}(2022)]%
        {rotaru_event-level_2022}
\bibfield{author}{\bibinfo{person}{Victor Rotaru}, \bibinfo{person}{Yi Huang}, \bibinfo{person}{Timmy Li}, \bibinfo{person}{James Evans}, {and} \bibinfo{person}{Ishanu Chattopadhyay}.} \bibinfo{year}{2022}\natexlab{}.
\newblock \showarticletitle{Event-level prediction of urban crime reveals a signature of enforcement bias in {US} cities}.
\newblock \bibinfo{journal}{\emph{Nature Human Behaviour}} \bibinfo{volume}{6}, \bibinfo{number}{8} (\bibinfo{date}{Aug.} \bibinfo{year}{2022}), \bibinfo{pages}{1056--1068}.
\newblock
\showISSN{2397-3374}
\urldef\tempurl%
\url{https://doi.org/10.1038/s41562-022-01372-0}
\showDOI{\tempurl}
\newblock
\shownote{Number: 8 Publisher: Nature Publishing Group}.


\bibitem[Roulston and Choi(2018)]%
        {roulston_sage_2018}
\bibfield{author}{\bibinfo{person}{Kathryn Roulston} {and} \bibinfo{person}{Myungweon Choi}.} \bibinfo{year}{2018}\natexlab{}.
\newblock \bibinfo{booktitle}{\emph{The {SAGE} {Handbook} of {Qualitative} {Data} {Collection}}}.
\newblock \bibinfo{publisher}{SAGE Publications Ltd}.
\newblock
\showISBNx{978-1-5264-1607-0}
\urldef\tempurl%
\url{https://doi.org/10.4135/9781526416070}
\showDOI{\tempurl}


\bibitem[Selbst(2017)]%
        {selbst2017disparate}
\bibfield{author}{\bibinfo{person}{Andrew~D Selbst}.} \bibinfo{year}{2017}\natexlab{}.
\newblock \showarticletitle{Disparate impact in big data policing}.
\newblock \bibinfo{journal}{\emph{Ga. L. Rev.}}  \bibinfo{volume}{52} (\bibinfo{year}{2017}), \bibinfo{pages}{109}.
\newblock


\bibitem[Shapiro(2019)]%
        {shapiro_predictive_2019}
\bibfield{author}{\bibinfo{person}{Aaron Shapiro}.} \bibinfo{year}{2019}\natexlab{}.
\newblock \showarticletitle{Predictive {Policing} for {Reform}? {Indeterminacy} and {Intervention} in {Big} {Data} {Policing}}.
\newblock \bibinfo{journal}{\emph{Surveillance \& Society}} \bibinfo{volume}{17}, \bibinfo{number}{3/4} (\bibinfo{date}{Sept.} \bibinfo{year}{2019}), \bibinfo{pages}{456--472}.
\newblock
\showISSN{1477-7487}
\urldef\tempurl%
\url{https://doi.org/10.24908/ss.v17i3/4.10410}
\showDOI{\tempurl}


\bibitem[Simkin(2023)]%
        {simkin_looking_2023}
\bibfield{author}{\bibinfo{person}{Maya Simkin}.} \bibinfo{year}{2023}\natexlab{}.
\newblock \bibinfo{title}{Looking {Forward} to the {End} of {Chicago}'s {Contract} with {ShotSpotter}}.
\newblock
\newblock
\urldef\tempurl%
\url{https://www.chicagoappleseed.org/2023/05/10/ending-chicago-contract-with-shotspotter/}
\showURL{%
\tempurl}


\bibitem[Small(2009)]%
        {small_how_2009}
\bibfield{author}{\bibinfo{person}{Mario~Luis Small}.} \bibinfo{year}{2009}\natexlab{}.
\newblock \showarticletitle{`{How} many cases do {I} need?': {On} science and the logic of case selection in field-based research}.
\newblock \bibinfo{journal}{\emph{Ethnography}} \bibinfo{volume}{10}, \bibinfo{number}{1} (\bibinfo{date}{March} \bibinfo{year}{2009}), \bibinfo{pages}{5--38}.
\newblock
\showISSN{1466-1381}
\urldef\tempurl%
\url{https://doi.org/10.1177/1466138108099586}
\showDOI{\tempurl}
\newblock
\shownote{Publisher: SAGE Publications}.


\bibitem[Spencer et~al\mbox{.}(2016)]%
        {spencer2016implicit}
\bibfield{author}{\bibinfo{person}{Katherine~B Spencer}, \bibinfo{person}{Amanda~K Charbonneau}, {and} \bibinfo{person}{Jack Glaser}.} \bibinfo{year}{2016}\natexlab{}.
\newblock \showarticletitle{Implicit bias and policing}.
\newblock \bibinfo{journal}{\emph{Social and Personality Psychology Compass}} \bibinfo{volume}{10}, \bibinfo{number}{1} (\bibinfo{year}{2016}), \bibinfo{pages}{50--63}.
\newblock


\bibitem[Stevenson(2018)]%
        {stevenson_assessing_2018}
\bibfield{author}{\bibinfo{person}{Megan~T. Stevenson}.} \bibinfo{year}{2018}\natexlab{}.
\newblock \showarticletitle{Assessing {Risk} {Assessment} in {Action}}.
\newblock \bibinfo{journal}{\emph{Minnesota Law Review}}  \bibinfo{volume}{103} (\bibinfo{year}{2018}), \bibinfo{pages}{303}.
\newblock
\urldef\tempurl%
\url{https://papers.ssrn.com/abstract=3016088}
\showURL{%
\tempurl}


\bibitem[Stroud(2021)]%
        {stroud_automated_2021}
\bibfield{author}{\bibinfo{person}{Matt Stroud}.} \bibinfo{year}{2021}\natexlab{}.
\newblock \bibinfo{title}{An automated policing program got this man shot twice}.
\newblock
\newblock
\urldef\tempurl%
\url{https://www.theverge.com/c/22444020/heat-listed-csk-entry}
\showURL{%
\tempurl}


\bibitem[Sweeney(2020)]%
        {sweeney_for_2020}
\bibfield{author}{\bibinfo{person}{Annie Sweeney}.} \bibinfo{year}{2020}\natexlab{}.
\newblock \bibinfo{title}{For years {Chicago} police rated the risk of tens of thousands being caught up in violence. {That} controversial effort has quietly been ended.}
\newblock
\newblock
\urldef\tempurl%
\url{https://www.chicagotribune.com/2020/01/24/for-years-chicago-police-rated-the-risk-of-tens-of-thousands-being-caught-up-in-violence-that-controversial-effort-has-quietly-been-ended/}
\showURL{%
\tempurl}


\bibitem[Umansky(2023)]%
        {umansky_failed_2023}
\bibfield{author}{\bibinfo{person}{Eric Umansky}.} \bibinfo{year}{2023}\natexlab{}.
\newblock \showarticletitle{The {Failed} {Promise} of {Police} {Body} {Cameras}}.
\newblock \bibinfo{journal}{\emph{The New York Times}} (\bibinfo{date}{Dec.} \bibinfo{year}{2023}).
\newblock
\showISSN{0362-4331}
\urldef\tempurl%
\url{https://www.nytimes.com/2023/12/13/magazine/police-body-cameras-miguel-richards.html}
\showURL{%
\tempurl}


\bibitem[Van~Craen and Skogan(2017)]%
        {van2017achieving}
\bibfield{author}{\bibinfo{person}{Maarten Van~Craen} {and} \bibinfo{person}{Wesley~G Skogan}.} \bibinfo{year}{2017}\natexlab{}.
\newblock \showarticletitle{Achieving fairness in policing: The link between internal and external procedural justice}.
\newblock \bibinfo{journal}{\emph{Police quarterly}} \bibinfo{volume}{20}, \bibinfo{number}{1} (\bibinfo{year}{2017}), \bibinfo{pages}{3--23}.
\newblock


\bibitem[Verma(2023)]%
        {verma_heat_2021}
\bibfield{author}{\bibinfo{person}{Pransha Verma}.} \bibinfo{year}{2023}\natexlab{}.
\newblock \showarticletitle{Heat {Listed}}.
\newblock \bibinfo{journal}{\emph{The Washington Post}} (\bibinfo{date}{Dec.} \bibinfo{year}{2023}).
\newblock
\showISSN{0362-4331}
\urldef\tempurl%
\url{https://www.washingtonpost.com/technology/2022/07/15/predictive-policing-algorithms-fail/}
\showURL{%
\tempurl}


\bibitem[Webster(2021)]%
        {webster2021if}
\bibfield{author}{\bibinfo{person}{Richard~A Webster}.} \bibinfo{year}{2021}\natexlab{}.
\newblock \showarticletitle{“If everybody’s white, there can’t be any racial bias”: The disappearance of hispanic drivers from traffic records}.
\newblock \bibinfo{journal}{\emph{ProPublica--Journalism in the Public Interest. https://www. propublica. org/article/if-everybodys-white-there-cant-be-any-racial-bias-the-disappearance-of-hispanic-drivers-from-traffic-records}} (\bibinfo{year}{2021}).
\newblock


\bibitem[Weiss(1994)]%
        {weiss_learning_1994}
\bibfield{author}{\bibinfo{person}{Robert~Stuart Weiss}.} \bibinfo{year}{1994}\natexlab{}.
\newblock \bibinfo{booktitle}{\emph{Learning from {Strangers}: {The} {Art} and {Method} of {Qualitative} {Interview} {Studies}}}.
\newblock \bibinfo{publisher}{Free Press}.
\newblock
\showISBNx{978-0-02-934625-9}
\newblock
\shownote{Google-Books-ID: SLm6AAAAIAAJ}.


\bibitem[White et~al\mbox{.}(2021)]%
        {white2021can}
\bibfield{author}{\bibinfo{person}{Michael~D White}, \bibinfo{person}{Henry~F Fradella}, {and} \bibinfo{person}{Michaela Flippin}.} \bibinfo{year}{2021}\natexlab{}.
\newblock \showarticletitle{How can we achieve accountability in policing? The (not-so-secret) ingredients to effective police reform}.
\newblock \bibinfo{journal}{\emph{Lewis \& Clark L. Rev.}}  \bibinfo{volume}{25} (\bibinfo{year}{2021}), \bibinfo{pages}{405}.
\newblock


\bibitem[Widder and Nafus(2023)]%
        {widder2023dislocated}
\bibfield{author}{\bibinfo{person}{David~Gray Widder} {and} \bibinfo{person}{Dawn Nafus}.} \bibinfo{year}{2023}\natexlab{}.
\newblock \showarticletitle{Dislocated accountabilities in the “AI supply chain”: Modularity and developers’ notions of responsibility}.
\newblock \bibinfo{journal}{\emph{Big Data \& Society}} \bibinfo{volume}{10}, \bibinfo{number}{1} (\bibinfo{year}{2023}), \bibinfo{pages}{20539517231177620}.
\newblock


\bibitem[Widder et~al\mbox{.}(2022)]%
        {widder2022limits}
\bibfield{author}{\bibinfo{person}{David~Gray Widder}, \bibinfo{person}{Dawn Nafus}, \bibinfo{person}{Laura Dabbish}, {and} \bibinfo{person}{James Herbsleb}.} \bibinfo{year}{2022}\natexlab{}.
\newblock \showarticletitle{Limits and possibilities for “Ethical AI” in open source: A study of deepfakes}. In \bibinfo{booktitle}{\emph{Proceedings of the 2022 ACM Conference on Fairness, Accountability, and Transparency}}. \bibinfo{pages}{2035--2046}.
\newblock


\bibitem[Yeung et~al\mbox{.}(2021)]%
        {yeung_identifying_2021}
\bibfield{author}{\bibinfo{person}{Douglas Yeung}, \bibinfo{person}{Inez Khan}, \bibinfo{person}{Nidhi Kalra}, {and} \bibinfo{person}{Osonde~A. Osoba}.} \bibinfo{year}{2021}\natexlab{}.
\newblock \bibinfo{booktitle}{\emph{Identifying {Systemic} {Bias} in the {Acquisition} of {Machine} {Learning} {Decision} {Aids} for {Law} {Enforcement} {Applications}}}.
\newblock \bibinfo{type}{{T}echnical {R}eport}. \bibinfo{institution}{RAND Corporation}.
\newblock
\urldef\tempurl%
\url{https://www.rand.org/pubs/perspectives/PEA862-1.html}
\showURL{%
\tempurl}


\bibitem[Zhang et~al\mbox{.}(2018)]%
        {zhang_mitigating_2018}
\bibfield{author}{\bibinfo{person}{Brian~Hu Zhang}, \bibinfo{person}{Blake Lemoine}, {and} \bibinfo{person}{Margaret Mitchell}.} \bibinfo{year}{2018}\natexlab{}.
\newblock \bibinfo{title}{Mitigating {Unwanted} {Biases} with {Adversarial} {Learning}}.
\newblock
\newblock
\urldef\tempurl%
\url{https://doi.org/10.48550/arXiv.1801.07593}
\showDOI{\tempurl}
\newblock
\shownote{arXiv:1801.07593 [cs]}.


\bibitem[Zilka et~al\mbox{.}(2023)]%
        {zilka_progression_2023}
\bibfield{author}{\bibinfo{person}{Miri Zilka}, \bibinfo{person}{Riccardo Fogliato}, \bibinfo{person}{Jiri Hron}, \bibinfo{person}{Bradley Butcher}, \bibinfo{person}{Carolyn Ashurst}, {and} \bibinfo{person}{Adrian Weller}.} \bibinfo{year}{2023}\natexlab{}.
\newblock \showarticletitle{The {Progression} of {Disparities} within the {Criminal} {Justice} {System}: {Differential} {Enforcement} and {Risk} {Assessment} {Instruments}}. In \bibinfo{booktitle}{\emph{Proceedings of the 2023 {ACM} {Conference} on {Fairness}, {Accountability}, and {Transparency}}} \emph{(\bibinfo{series}{{FAccT} '23})}. \bibinfo{publisher}{Association for Computing Machinery}, \bibinfo{address}{New York, NY, USA}, \bibinfo{pages}{1553--1569}.
\newblock
\showISBNx{9798400701924}
\urldef\tempurl%
\url{https://doi.org/10.1145/3593013.3594099}
\showDOI{\tempurl}


\end{thebibliography}

\appendix
\newpage 

\section{Interview Guide}
\label{sec:interview_guide}

\noindent \begin{enumerate}
	\item \textbf{Context and Positionality} 
	
	\begin{itemize}
		\item Can you introduce yourself? 
        \item Can you tell us more about your work?
        \item How are you related to the predictive policing project? How have you heard of it, if at all?
        \item How are you involved with other initiatives around community and public safety in Chicago?
	\end{itemize}

	\item \textbf{Bias definition and utility}

	
	\begin{itemize}
		\item What is algorithmic bias to you in general? 
   \item What is the problem with it? Why do you think it exists?
        \item What is enforcement bias to you specifically?
        \item What is the problem with it? Why do you think it exists?
        \item What is its specific relevance to your work? And the context of Chicago?
        \item What is your role in tackling bias and problems faced in doing so (if applicable) at work?
		
	\end{itemize}

	\item \textbf{Perceived feasibility and desirability of using evidence of bias to tackle social disparities} 
	
 
	\begin{itemize}
		\item How should bias be addressed? 
        \item What would an ideal solution look like to you (especially with relevance to your role/work)?
        \item What are the barriers to resolving bias? How can it be realistically tackled, in your opinion?
	\end{itemize}

	\item \textbf{Snowball sampling}

	\begin{itemize}
		\item Do you know other people it might be relevant for us to interview?
		
	\end{itemize}
	
\end{enumerate}



\section{Code Table}
\label{sec:code_table}

\small
\begin{supertabular}{  l } 
 
	\textbf{Abolition} \\ 
	
	\hline
	
    Alternatives to current system/police\\
    Building healthy communities\\
    Calls for changing the narrative/status quo\\
    The need to acknowledge harms of current system\\
    Importance of history and education\\
    Importance of hope and belief in change\vspace{.2em}\\

	\textbf{Co-option and exploitation} 		\\	\hline

    Academic exploitation\\
    Co-opting evidence-based label\\
    Commercial interest in predicting crime\\
    Disruption rhetoric \\
    Open source as a marketing strategy\\
    Responsibilities of developers\vspace{.2em}\\

	\textbf{Downsides of AI}	\\
	
	\hline

    AI is racist/not objective\\
    Inherent reservations about AI\\
    Garbage in garbage out/distrust of data\\
    Human discretion is important\\
    Nuance not captured by algorithm\\
    Technology isn't going to fix policing\vspace{.2em}\\
	
	\textbf{Lived experience and positionality}	\\		\hline

    Affected people are best suited to solve problems\\
    Barriers to entering research\\
    Impacted groups lack resources\\
    Different stakeholders care about different forms of bias\\
    People not affected don't have the same understanding\\
    Who generated the data/designed the tool matters\\
    Importance of trust\vspace{.2em}\\

	\textbf{Reformism}\\			\hline

    Interpersonal bias rather than structural\\
    Resource allocation\\
    Safety is the priority\\
    Police promotes safety\vspace{.2em}\\

	\textbf{Structural bias}\\
	\hline
    Importance of addressing root cause of crime\\
    Racial profiling and segregation\\
    Structural racism requires a structural solution\\
    The system is biased\vspace{.2em}\\
	
	\textbf{Upsides of AI}\\
	\hline

    Algorithm can be more objective/humans are biased\\
    Algorithms increase efficiency \\
    Open source\\
    Using data to understand authority figures (sousveillance)\\
    Information has multiple uses\vspace{.2em}\\
	
	\hline
\end{supertabular}


\end{document}